\documentclass[iop, numberedappendix]{mn2e}

\usepackage[pdftex]{graphicx}

\usepackage{amsmath, amsthm, amssymb}
\let\bibhang\relax
\usepackage{natbib}
\usepackage{subfigure}
\usepackage{upgreek}
\usepackage{accents}
\usepackage{color}
\usepackage{hyperref}
\usepackage{dblfloatfix}
\usepackage[T1]{fontenc}
\usepackage{aecompl}
\usepackage{enumitem}
\usepackage{pifont}
\newcommand{\cmark}{\ding{51}}%
\newcommand{\xmark}{\ding{55}}%

\newcommand{\angstrom}{\text{\normalfont\AA}}

\hypersetup{
  colorlinks   = true, 
  urlcolor     = blue, 
  linkcolor    = blue, 
  citecolor   = blue 
}

\newcommand{\AUV}{A_{\text{UV}}}

\newcommand{\Rvir}{R_{\text{vir}}}

\newcommand{\sigmaN}{\sigma_{\log_{10} N_d}}

\newcommand{\xLAE}{x_{\text{LAE}}}

\newcommand{\zpr}{z^{\prime}}

\newcommand{\obs}{\text{obs}}

\newcommand{\zobs}{z_{\obs}}
\newcommand{\vir}{\text{vir}}

\newcommand{\Lya}{\text{Ly-}\alpha}

\newcommand{\alphalo}{\alpha_{\ast,\text{lo}}}
\newcommand{\alphahi}{\alpha_{\ast,\text{hi}}}

\newcommand{\alphaloD}{\alpha_{d,\text{lo}}}
\newcommand{\alphahiD}{\alpha_{d,\text{hi}}}

\newcommand{\MUV}{M_{\text{UV}}}
\newcommand{\MUVobs}{M_{\text{UV,obs}}}

\newcommand{\Mstell}{M_{\ast}}
\newcommand{\Mpeak}{M_{\ast,\text{peak}}}

\newcommand{\SFR}{\dot{M}_{\ast}}

\newcommand{\fstar}{f_{\ast}}

\newcommand{\Msun}{M_{\odot}}

\newcommand{\zform}{z_{\text{form}}}

\title[Rest UV colours of high-$z$ galaxies]{Effects of self-consistent rest-ultraviolet colours in semi-empirical galaxy formation models}
\author[Mirocha, Mason, \& Stark]{
Jordan Mirocha,$^{1}$\textsuperscript{\thanks{jordan.mirocha@mcgill.ca}}\textsuperscript{\thanks{CITA National Fellow}}
Charlotte Mason,$^{2}$\textsuperscript{\thanks{Hubble Fellow}} and
Daniel P. Stark$^{3}$  \\
$^{1}$McGill University Department of Physics \& McGill Space Institute, 3600 Rue University, Montr\'eal, QC, H3A 2T8 Canada \\
$^{2}$Center for Astrophysics | Harvard \& Smithsonian, 60 Garden St, Cambridge, MA, 02138, USA \\
$^{3}$Steward Observatory, University of Arizona, 933 N Cherry Ave, Tucson, AZ 85721 USA
}

\begin{document}

\pagerange{\pageref{firstpage}--\pageref{lastpage}} \pubyear{2016}
\maketitle

\begin{abstract}
Connecting the observed rest-ultraviolet (UV) luminosities of high-$z$ galaxies to their intrinsic luminosities (and thus star formation rates) requires correcting for the presence of dust. We bypass a common dust-correction approach that uses empirical relationships between infrared (IR) emission and UV colours, and instead augment a semi-empirical model for galaxy formation with a simple -- but self-consistent -- dust model and use it to jointly fit high-$z$ rest-UV luminosity functions (LFs) and colour-magnitude relations ($\MUV$-$\beta$). In doing so, we find that UV colours evolve with redshift (at fixed UV magnitude), as suggested by observations, even in cases without underlying evolution in dust production, destruction, absorption, or geometry. The observed evolution in our model arises due to the reduction in the mean stellar age and rise in specific star formation rates with increasing $z$. The UV extinction, $\AUV$, evolves similarly with redshift, though we find a systematically shallower relation between $\AUV$ and $\MUV$ than that predicted by IRX-$\beta$ relationships derived from $z \sim 3$ galaxy samples. Finally, assuming that high $1600\angstrom$ transmission ($\gtrsim 0.6$) is a reliable LAE indicator, modest scatter in the effective dust surface density of galaxies can explain the evolution both in $\MUV$-$\beta$ and LAE fractions. These predictions are readily testable by deep surveys with the \textit{James Webb Space Telescope}.
\end{abstract}
\begin{keywords}
galaxies: high-redshift -- galaxies: luminosity function, mass function.
\end{keywords}
%


\section{Introduction}
Current constraints on galaxy formation are based largely on the rest ultraviolet (UV) properties of redshift $z \gtrsim 4$ galaxies, e.g., luminosity functions \citep[LFs;][]{Bouwens2015,Finkelstein2015} and UV colour-magnitude relations \citep[CMDs, $\MUV$-$\beta$;][]{Bouwens2009,Finkelstein2012,Bouwens2014,Dunlop2013}. Such observations probe the star formation rate (SFR) of high-$z$ galaxies, given that the rest-UV emission is dominated by massive young stars, and have thus allowed astronomers to begin piecing together the cosmic star formation rate density (SFRD) in the early Universe \citep[see][for a review]{Madau2014}. UV colours, generally quantified by a power-law spectral slope $\beta$ (defined by $f_{\lambda} \propto \lambda^{\beta}$), are critical to this inference as they are modulated by dust extinction in a characteristic wavelength-dependent manner, allowing one to ``dust correct'' UV magnitude measurements, so long as multi-band photometry covering the rest UV continuum ($1300 \lesssim \lambda/\angstrom \lesssim 2600$) is available.

Unfortunately, the link between $\beta$ and UV extinction, $\AUV$, is potentially complicated. One common approach is to assume that thermal radiation emitted from dust grains (in the infrared; IR) is a reliable tracer of the energy lost in the UV. In low redshift star-forming galaxies, for which there is both rest-UV and rest-IR coverage, there is indeed a relationship between a galaxy's infrared ``excess'' and its UV slope \citep[the so-called IRX-$\beta$ relationship;][]{Meurer1999}. For a known input stellar spectrum, assumed dust opacity (as a function of wavelength), and bolometric correction (to recover total dust emission from narrow-band IR observations), one can determine $\AUV$ from $\beta$, and thus convert observed magnitudes to intrinsic magnitudes. \citet{Meurer1999} (hereafter M99) found that $A_{1600} = 4.43 - 1.99 \beta$, using a \citet{Calzetti1994} (hereafter C94) dust law and input stellar spectra from \textsc{starburst99} \citep{Leitherer1999}.

\defcitealias{Meurer1999}{M99}
\defcitealias{Calzetti1994}{C94}

In recent years, many groups have adopted some variant of the IRX-$\beta$-based procedure as a way to use observed UVLFs and CMDs to calibrate semi-analytic models (SAMs), which may not model dust explicitly. There are several ways this approach may break down. For example, the origin of the IRX-$\beta$ relation is an active area of research, both observationally \citep[e.g.,][]{Overzier2011,Casey2014,Reddy2018} and theoretically \citep[e.g.,][]{Narayanan2018,Salim2019,Ma2019,Schulz2020}, so it may be premature to apply it at arbitrarily high redshift, where the properties of stars and dust may differ from low-$z$ samples. Inferences based on IRX-$\beta$ arguments could be biased for a less interesting reason, which is that the assumptions underlying the \citetalias{Meurer1999} relation are often not made in SAMs. For example, adopting a different stellar population synthesis (SPS) model can change the input stellar spectrum, as can changes in the star formation histories (SFHs) of individual galaxies, thus posing a self-consistency issue when invoking empirical IRX-$\beta$ relations. For example, model galaxies generally have rapidly rising star formation histories (SFHs), and are thus intrinsically bluer than the assumed input $\beta$ adopted in \citetalias{Meurer1999} \citep[see, e.g.,][]{Finlator2011,Wilkins2013,Mancini2016}, which is based on the assumption of a constant star formation rate (SFR).

In this work we take a different approach that avoids self-consistency issues by using a simple dust model in lieu of an IRX-$\beta$ assumption. This approach allows us to compute self-consistent solutions for the spectra of objects in our model, assess the circumstances in which evolution in the properties of dust is required by current measurements, and make physically-motivated predictions for upcoming $\beta$ measurements to be conducted with the \textit{James Webb Space Telescope} (JWST).


In Section 2 we detail our dust model and the underlying assumptions about star formation in early galaxies. We detail our main results in Section 3 and discuss them in a broader context in Section 4. In Section 5 we summarize our findings.

We adopt AB magnitudes throughout \citep{Oke1983}, i.e.,
\begin{equation}
    M_{\lambda} = -2.5 \log_{10} \left(\frac{f_{\lambda}}{3631 \ \mathrm{Jy}} \right)
\end{equation}
and adopt the following cosmology: $\Omega_m = 0.3156$, $\Omega_b = 0.0491$, $h = 0.6726$, and $\sigma_8=0.8159$, very similar to the recent \citet{Planck2018} constraints.

\section{Model}
Our model is similar to other semi-empirical models that have appeared in the literature in recent years. We outline our model for star formation in galaxies in \S\ref{sec:stars}, our approach to dust in \S\ref{sec:dust}, and our method for generating synthetic spectra and estimating UV colours in \S\ref{sec:synthobs}. Much of this has been described previously \citep{Mirocha2017,Mirocha2019}, and is publicly available within the \textsc{ares}\footnote{\url{https://ares.readthedocs.io/en/latest/}} code.

\subsection{Star Formation} \label{sec:stars}
We assume the SFR is proportional to the baryonic mass accretion (MAR) onto dark matter (DM) halos \citep[as in, e.g.;][]{Mason2015,Sun2016} i.e.,
\begin{equation}
    \dot{M}_{\ast}(M_h, z) = \fstar(M_h, z) \dot{M}_b(z, M_h) \label{eq:SFR}
\end{equation}
where $\fstar$ is the efficiency of star formation. The baryonic MAR is well approximated by a power-law in mass and redshift \citep[e.g.,][]{McBride2009,Dekel2013}, however, rather than adopting a parametric form for the MAR calibrated by simulations, we derive it directly from the halo mass function \citep[HMF; see Appendix A of][for more details]{Furlanetto2017}. We adopt the \citet{Tinker2010} HMF in this work, generated by the \textsc{hmf} code\footnote{\url{https://hmf.readthedocs.io/en/latest/}} \citep{Murray2013}.

We assume that the star formation efficiency (SFE) is a double power-law in $M_h$ \citep{Moster2010}, i.e.,
\begin{equation}
    \fstar(M_h) = \frac{f_{\ast,10} \ \mathcal{C}_{10}} {\left(\frac{M_h}{M_{\mathrm{p}}} \right)^{-\alphalo} + \left(\frac{M_h}{M_{\mathrm{p}}} \right)^{-\alphahi}} \label{eq:sfe_dpl}
\end{equation}
where $f_{\ast,10}$ is the SFE at $10^{10} M_{\odot}$, $M_p$ is the mass at which $\fstar$ peaks, and $\alphahi$ and $\alphalo$ describe the power-law index at masses above and below the peak, respectively. The additional constant $\mathcal{C}_{10} \equiv (10^{10} / M_p)^{-\alphalo} + (10^{10} / M_p)^{-\alphahi}$ is introduced to re-normalize the standard double power-law formula to $10^{10} M_{\odot}$, rather than the peak mass. This model predicts $z > 6$ UVLFs in good agreement with observations when calibrating only to measurements at $z\sim 6$ \citep{Mirocha2017,Furlanetto2017}, in agreement with the results of other similar models from recent studies \citep[e.g.,][]{Trenti2010,Behroozi2013a,Mason2015,Mashian2016,Sun2016,Tacchella2018,Behroozi2019}.

In this work, we extend our models to $z \sim 4$ to more adequately address issues of time evolution in the $\MUV$-$\beta$ relationship and UVLFs. This redshift range, though somewhat arbitrary, has become the de-facto interval for this kind of modeling in recent years, presumably due to the availabity of homogenous datasets, though pushing to even lower redshifts would of coure be advantageous and is a current work in progress. We include log-normal scatter in the SFR of halos at fixed halo mass, $\sigma_{\log_{10} \mathrm{SFR}} = 0.3$, which we take to represent scatter in halo accretion rates, but such scatter will also resemble the dispersion in halo assembly times \citep[e.g.,][]{Ren2018}. We then synthesize the spectra of all galaxies in the model, rather than assuming a constant steady-state value for the relationship between UV luminosity and SFR. In other words, for each halo in our model, with index $i$, we determine the intrinsic spectrum at redshift $z_{\obs}$ by integrating over the past star formation history, i.e.,
\begin{equation}
    L_{\lambda,i} (\zobs) = \int_{\zobs}^{\zform} \dot{M}_{\ast,i}(\zpr) l_{\lambda}(\Delta t^{\prime}) \bigg|\frac{dt}{dz} \bigg| dz
\end{equation}
where $l_{\lambda}(\Delta t^{\prime})$ is the luminosity of a simple stellar population of age $\Delta t = t(\zobs) - t(\zpr)$. We include both the stellar continuum, provided by \textsc{bpass}, and nebular continuum emission following standard procedures: for free-free and free-bound emission, we use the emission coefficients of \citet{Ferland1980} (his Table 1), while for the two-photon emission probability we take the fitting formula of \citet{Fernandez2006} (their Eq. 22), which was derived from the tabulated results of \citet{Brown1970}. We adopt a temperature of $2 \times 10^4$ K in HII regions when generating the nebular continuum. We find that inclusion of the nebular continuum constitutes a small correction, equivalent to a $\sim 10$\% reduction in dust content.

The synthesized luminosity is then reddened by an optical depth $\tau_{\lambda,i}$, yielding the observed luminosity
\begin{equation}
    L^{\prime}_{\lambda,i} = L_{\lambda,i} (\zobs) e^{-\tau_{\lambda,i}} \label{eq:reddening}
\end{equation}
We adopt the \textsc{bpass} version 1.0 \citep{Eldridge2009} single-star models throughout when modeling $l_{\lambda}(\Delta t^{\prime})$ with an intermediate stellar metallicity of $Z_{\odot}/5$. These choices largely affect the inferred normalization of the SFE and dust opacity. As a result, any change to the stellar model will largely be absorbed by normalization parameters, leaving constraints on the \textit{shape} of the SFE and dust scale length relatively unaffected.

\subsection{Dust Obscuration} \label{sec:dust}
For each galaxy in our model we track the build-up of metals by assuming a fixed metal yield per unit SFR, i.e., $\dot{M}_Z = f_Z \dot{M}_{\ast}$, where the metal production efficiency $f_Z$ is set to 0.1 in our fiducial case. We further assume that a fraction $f_d=0.4$ of these metals reside in dust grains \citep{Dwek1998}. This ``instantaneous recyling'' approximation is reasonable, at least in the $z \gtrsim 6$ limit, as the Universe is too young for asymptotic giant branch stars to have become a non-negligible source of dust production \citep[e.g.,][]{Dwek2007}.

To redden galaxy spectra, we must also make an assumption about the geometry of the dust distribution and the opacity of dust (per unit mass). For simplicity, we adopt a simple spherically-symmetric dust screen model, where the dust optical depth along obscured lines of sight is given by
\begin{equation}
    \tau_{\lambda} = \int_0^R \rho_{\mathrm{dust}}(r) \kappa_{\lambda} dr .
\end{equation}
We take the absorption coefficient $\kappa$ to be a power-law as we only explore a relatively narrow range in wavelength in this study,
\begin{equation}
    \kappa_{\lambda} = \kappa_{1000} \left(\frac{\lambda}{10^3 \angstrom} \right)^{\gamma_{\kappa}} \label{eq:kappa}
\end{equation}
where $\kappa_{1000} = \kappa(\lambda=10^3\angstrom) \equiv 10^5 \mathrm{cm}^{2} \ \mathrm{g}^{-1} \simeq 20 \ \mathrm{pc}^2 M_{\odot}^{-1}$ and $\gamma_{\kappa} \equiv -1$ in our fiducial model. These choices are consistent with an SMC-like dust law in the rest-UV \citep{Weingartner2001}, though more complex models may be warranted, e.g., to accommodate the 2175\angstrom\ ``bump'' present in some galaxies \citep[for a recent review of the dust attenuation law, see][]{Salim2020}.

In this framework, sources at the center of spherically-symmetric, uniform density dust clouds, are obscured by an optical depth given by
\begin{equation}
    \tau_{\lambda} = \kappa_{\lambda} N_d = \kappa_{\lambda} \frac{3 M_d}{4 \pi R_d^2} \label{eq:tau_d}
\end{equation}
i.e., the characteristic scale $R_d$ determines both the dust density and the length of sightlines passing through the dust envelope. To start, we model $R_d$ generically as a power-law in mass,
\begin{equation}
    R_d = R_0 \left(\frac{M}{10^{10} M_{\odot}} \right)^{\alpha_{d}} \ \mathrm{kpc},
\end{equation}
where $R_0$ normalizes the scale length at $M_h=10^{10}\ \Msun$ and $\alpha_d$ controls the dependence of $R_d$ on $M_h$. Similar approaches have been taken in previous work, e.g., \citet{Somerville2012} adopt $R_d = R_{\mathrm{gas}}$, where the radius of the cold gas disk $R_{\mathrm{gas}}$ is assumed to be a constant fraction of the stellar scale length. Note that the virial radii of dark matter halos evolve as $R_{\vir} \propto M_h^{1/3} (1+z)^{-1}$, while $R_d \propto M_h^{1/2}$ implies dust column densities that are proportional to $M_d / M_h$ (see Eq. \ref{eq:tau_d}).

We will show in \S\ref{sec:results} that a more complicated function is likely warranted, at which point we will employ a double power-law for $R_d$ as well as $f_{\ast}$, i.e.,
\begin{equation}
    R_d(M_h) = \frac{R_{d,10} \ \mathcal{D}_{10}} {\left(\frac{M_h}{M_{\mathrm{p}}} \right)^{-\alphaloD} + \left(\frac{M_h}{M_{\mathrm{p}}} \right)^{-\alphahiD}} \label{eq:Rd_dpl}
\end{equation}
where each parameter is analogous to those in Eq. \ref{eq:sfe_dpl}. We discuss our choice of $R_d$ parameterization further in \S\ref{sec:calib} and \S\ref{sec:evolution}.

We make no effort to model dust \textit{emission} in this work -- any link to the IRX-$\beta$ relation would require modeling of dust temperatures. \citet{Imara2018} present a very similar approach to ours, but focus instead on the implications at longer wavelengths. We expect that our predictions for dust emission would be similar to those of \citet{Imara2018} were we to make the same assumptions for how stellar radiation is reprocessed by dust. As we will discuss in \S\ref{sec:discussion}, UV extinction is a prediction of our model, rather than an input, as is effectively the case for IRX-$\beta$-based models.

\subsection{Synthetic Observations} \label{sec:synthobs}
In order to fairly compare with constraints on the $\MUV$-$\beta$ relation at high-$z$, we ``observe'' our model galaxies using the same magnitude definition and photometric filters as in \citet{Bouwens2014} (hereafter B14). The filters\footnote{WFC:\url{http://www.stsci.edu/hst/acs/analysis/throughputs/tables}}\footnote{WFC3:\url{http://www.stsci.edu/hst/wfc3/ins\_performance/throughputs/Throughput\_Tables}} employed vary with redshift as follows:
\begin{itemize}
	\item $z\sim 4$: $i_{775}$, $I_{814}$, $z_{850}$, ($Y_{105}$), $J_{125}$
	\item $z\sim 5$: $z_{850}$, $Y_{105}$, ($J_{125}$), $H_{160}$
	\item $z\sim 6$: $Y_{105}$, ($J_{125}$), $H_{160}$
	\item $z\sim 7$: $J_{125}$, $H_{160}$
\end{itemize}

\defcitealias{Bouwens2014}{B14}

\citetalias{Bouwens2014} employed the filters listed above in the ERS \citep{Windhorst2011} and  CANDELS \citep{Koekemoer2011,Grogin2011} fields, but in the deeper XDF \citep{Illingworth2013}, and HUDF09 \citep{Bouwens2011} fields, filters enclosed in parentheses were not used. The $Y_{098}$ filter was used when available. The UV magnitude in \citetalias{Bouwens2014} is defined as the geometric mean of the photometric measurements for each galaxy, which we indicate with angular brackets, $\langle \MUV \rangle$.

We make no effort in this study to conduct mock surveys and perform sample selection self-consistently, and therefore have no basis on which to use different combinations of filters for objects at the same redshift. For consistency, we adopt the ERS/CANDELS filters for all objects, and only use the $Y_{098}$ filter at $4 \lesssim z \lesssim 6$. We expect this to be a reasonable approach given that most of the information about dust is in the brightest, reddest objects, which are captured best by the wider field surveys.

Our UV slope predictions are based on two slightly different approaches: (i) an empirically-motivated approach, in which $\beta$ is defined as a power-law fit through the available photometry, as in \citetalias{Bouwens2014}, and (ii) a more theoretical approach, in which $\beta$ is defined as a power-law fit to the ``true'' galaxy spectrum, i.e., the spectrum generated by our forward model (see Eqs. \ref{eq:sfe_dpl}-\ref{eq:reddening}) at its native resolution, sampled by the \citetalias{Calzetti1994} windows, which we indicate as $\beta_{\mathrm{c94}}$, as in \citet{Finkelstein2012}.

As pointed out in \citet{Finkelstein2012}, photometric estimates of $\beta$ can bias estimates of dust attenuation. This is largely for two reasons: (i) the spectrum of galaxies is not a pure power-law, so the inferred UV slope can change because filters intersect different parts of the the rest UV spectrum for galaxies at slightly different redshifts, and (ii) absorption lines, particularly at the shortest wavelengths, are unavoidable with photometry, and thus suppress inferred magnitudes despite being unrelated to dust. High resolution spectra can circumvent these problems, simply by only including ``clean'' spectral windows in the fit (e.g., the \citetalias{Calzetti1994} windows, which exclude wavelength ranges with strong absorption/emission features). Alternatively, one can perform SED-fitting on photometric measurements, and estimate $\beta$ from the best-fit SED, rather than the photometry. For the duration of this paper, we adopt the purely empirical approach, both to remain consistent with \citetalias{Bouwens2014} and because it is computationally more efficient. In select cases, we compare to the $\beta_{\mathrm{c94}}$ approach and \citet{Finkelstein2012} measurements, generally finding good agreement.

We compare $\beta$ estimation techniques thoroughly in Appendix \ref{sec:dust_biases}, where we also provide a full listing of the HST filters and the redshifts at which they are used (Table \ref{tab:photometry}; shown graphically in Figure \ref{fig:photometry}) along with the NIRCAM wide and medium filters that lie within the range of the \citet{Calzetti1994} windows. Unless indicated otherwise, these are the filters used for $\beta$ estimation throughout this paper. We confirm that photometrically-estimated UV slopes are biased relative to intrinsic, spectroscopically-estimated (in \citetalias{Calzetti1994} windows) UV slopes. The effect is generally of order $\Delta \beta \simeq +0.1$ for objects with intrinsic slopes $\beta \sim -2.4$, particularly at $z \gtrsim 6$ (for both HST and JWST), i.e., measured slopes are biased red. Note that we neglect other possible causes of biases in $\beta$ estimation, e.g., selection effects \citep[see, e.g.,][]{Dunlop2012}.

\section{Results} \label{sec:results}

\subsection{Basic Trends} \label{sec:basics}
\begin{figure*}
\begin{center}
\includegraphics[width=0.98\textwidth]{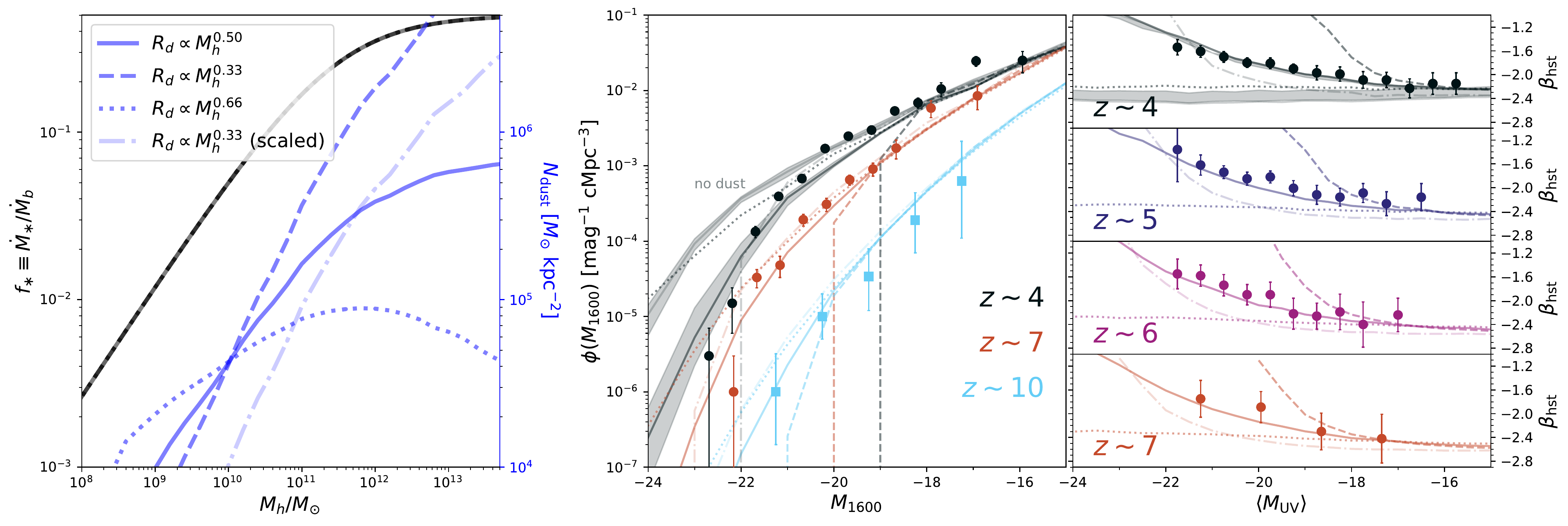}
\caption{{\bf Effects of variations in the relationship between dust scale length, $R_d$, and halo mass, $M_h$.} \textit{Left:} SFE (black) and dust column density (blue; right axis) as a function of $M_h$ for four different $R_d(M_h)$ models. Note that the dot-dashed curve is systematically shifted, but keeps the same power-law as the dashed curve. \textit{Right:} Corresponding UVLFs (left) and CMDs (right) at a series of redshifts. Models in which $R_d \propto M_h^{1/2}$ (solid) are in good agreement with measurements. Shallower slopes $R_d \propto M_h^{1/3}$ result in much too steep $\MUV-\beta$ relations (dashed), even if the normalization of $R_d$ is adjusted to systematically reduce dust reddening (dot-dashed). The shaded regions for $z\sim 4$ models represent mild, $d\kappa/d\lambda=\pm 0.3$ changes in the wavelength-dependence of the dust opacity (top-most bands), and differences caused by stellar metallicity for dust-free models (between $Z=0.001$ and $Z=Z_{\odot}=0.02$; horizontal bands). Data shown include UVLFs from \citet{Bouwens2015} ($4 \lesssim z \lesssim 8$) and \citet{Oesch2018} ($z \sim 10$), and $\MUV$-$\beta$ measurements from \citetalias{Bouwens2014}. A dust-free model is also shown for reference at $z \sim 4$. Note that these models are for illustrative purposes (i.e., they are not the result of fits; see \S\ref{sec:predictions} for MCMC results), and take $f_{\ast,10}=0.05$, $\Mpeak=2.8\times 10^{11} \ \Msun$, $\alphalo=2/3$, $\alphahi=0$, and $R_{d,10} = 1.4 \ \mathrm{kpc}$.}
\label{fig:model_dep_Rd}
\end{center}
\end{figure*}

In Figure \ref{fig:model_dep_Rd}, we show how the relationship between dust scale length, $R_d$, and halo mass, $M_h$, influences UVLFs and the $\MUV-\beta$ relation. We show scenarios spanning the range from $R_d \propto M_h^{1/3}$ to $R_d \propto M_h^{2/3}$.

For illustrative purposes, we fix $f_{\ast,10}=0.05$, $\Mpeak=2.8\times 10^{11} \ \Msun$, $\alphalo=2/3$, $\alphahi=0$, and $R_{d,10} = 1.4 \ \mathrm{kpc}$. In this case, the production of dust continues even as star formation slows at high mass, resulting in monotonically rising $\beta$. The \citetalias{Bouwens2014} measurements prefer $\alpha_d \simeq 0.5$. The UV colors are extremely sensitive to $\alpha_d$: the $\MUV-\beta$ relation becomes too steep for $\alpha_d = 1/3$ (dashed), and much too shallow for $\alpha_d = 2/3$ (dotted). In the former case, while reducing the normalization length scale, $R_{d,10}$, can help (equivalent to decrease in dust yield), the shape of the UVLF and $\MUV-\beta$ remain problematic (dash-dotted). The shaded region for $z\sim 4$ models shows how changing the wavelength-dependence of the dust opacity (Eq. \ref{eq:kappa}) between $\kappa_{\lambda} \propto \lambda^{-1.3}$ and $\kappa_{\lambda} \propto \lambda^{-0.7}$ affects the UVLFs and CMDs.

The sharp decline in the number counts of galaxies at the bright-end of the UVLF is generally interpreted to be in part a sign of dust reddening, but also a byproduct of a decline in the efficiency of star formation in high-mass halos \citep[e.g.,][]{Moster2010,Behroozi2010}. As a result, the assumption of a constant high-mass SFE used in Fig. \ref{fig:model_dep_Rd} is likely unreasonable. This complicates the simple power-law $R_d$ model used thus far, because the decline in the SFE also causes a decline in dust production, which, if sharp enough, can cause UV colours to start becoming \textit{bluer} as objects become brighter. Observations at $z \lesssim 4$ suggest that $\beta$ continues to rise monotonically for increasingly bright galaxies \citep[e.g.,][]{Lee2012}. In our model, ensuring that $\beta$ rises monotonically with decreasing $\MUV$ requires a change in how $R_d$ scales with halo mass.

\begin{figure*}
\begin{center}
\includegraphics[width=0.98\textwidth]{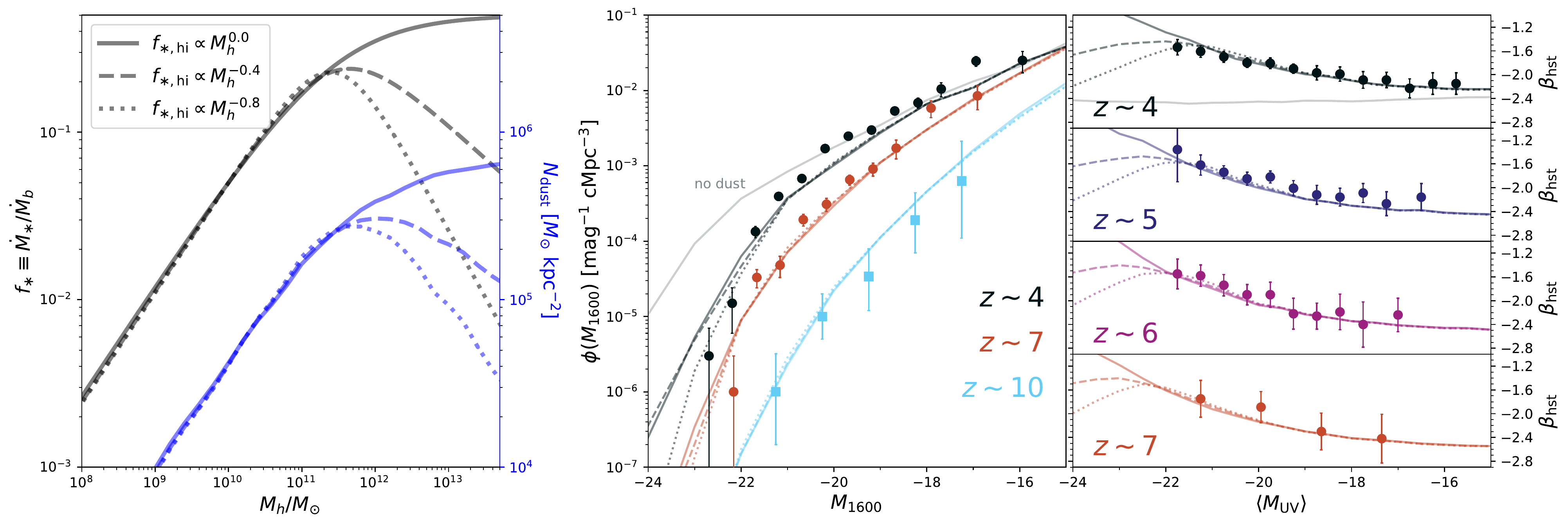}
\caption{{\bf Effects of variations in the efficiency of star formation in high-mass halos.} Same as Figure \ref{fig:model_dep_Rd}, except linestyles indicate a change in the slope of the SFE above the peak, $\alphahi$. Because dust production is directly proportional to galaxy SFR in our model, a downturn in the SFE can cause galaxies to become bluer as they grow more massive (dotted lines). Solutions to this issue are discussed in \S\ref{sec:calib}. Each model here takes $f_{\ast,10}=0.05$, $\Mpeak=2.8\times 10^{11} \ \Msun$, $\alphalo=2/3$, $R_{d,10} = 1.4 \ \mathrm{kpc}$, and $R_d \propto M_h^{1/2}$.}
\label{fig:model_dep_sfe}
\end{center}
\end{figure*}

We explore the impact of high-mass SFE variations in Figure \ref{fig:model_dep_sfe} for three different high-mass SFE slopes. For a strong decline, $\alphahi = -0.8$, predictions for the bright-end of the $z\sim 4$ UVLF are in much better agreement with \citetalias{Bouwens2014} measurements, though come at the cost of imparting a turn-over in $\MUV$-$\beta$ (dotted lines). This sharp decline in the SFE also has implications for galaxy stellar mass functions (SMFs), which we explore further in \S\ref{sec:calib}.

\subsection{Effects of Scatter in Dust Column Density} \label{sec:scatter}
The models shown thus far assume a 1:1 mapping between halo mass and dust scale length. Some scatter in the dust column density, $N_d$, at fixed $M_h$ still arises due to scatter in the SFR of galaxies (and thus dust production rate), but this is likely overly conservative. To explore the impact of scatter further we explore scenarios with lognormal scatter in $N_d$, $\sigmaN$, at fixed halo mass, deferring a discussion of intrinsic scatter in the $\MUV$-$\beta$ relation in \S\ref{sec:predictions}. In what follows, we also force the dust scale length to be a shallow function of halo mass, $R_d \propto M_h^{1/3}$, for reasons that will become apparent momentarily.

The introduction of $N_d$ scatter has an interesting impact on $\MUV-\beta$. Consider a faint galaxy, $\MUV \sim -17$, with the average amount of dust attenuation, so that $\beta \sim -2.4$. Now, if we subject this galaxy to a strong negative fluctuation in $N_d$, it will become brighter and bluer, and thus enter an $\MUV$ bin occupied (generally) by galaxies that reside in more massive, slightly more rare, halos. The opposite case of a positive $N_d$ spike will lead our galaxy to migrate in the opposite direction in the $\MUV-\beta$ plane, where it will occupy an $\MUV$ bin with galaxies that live in smaller, more common halos. As a result, there will be a net blueward bias in $\beta$ at fixed $\MUV$: galaxies scattering toward smaller $\MUV$ will always be outnumbered by unscattered objects in the same magnitude bin, while galaxies scattering to brighter $\MUV$ will always outnumber the ``typical'' galaxy in that bin. Note that this effect is strongest in models with the steepest UVLFs, and could thus be subject to revision if future observations find shallower UVLFs.

\begin{figure*}
\begin{center}
\includegraphics[width=0.98\textwidth]{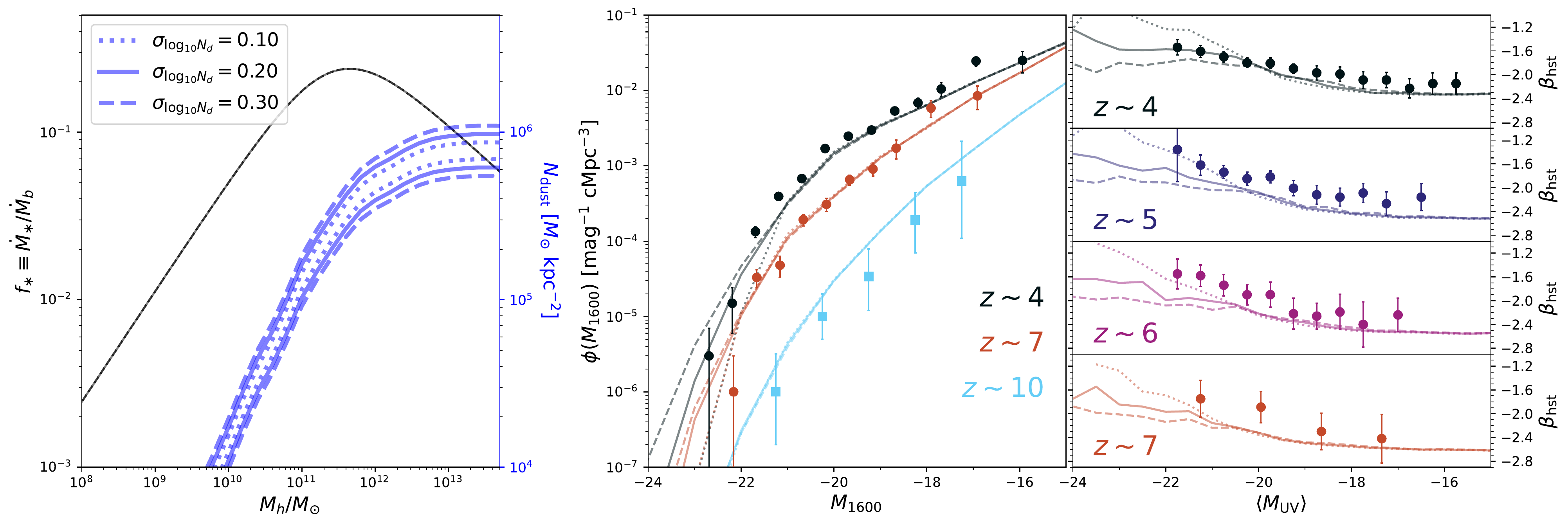}
\caption{{\bf Effects of scatter in dust column density at fixed halo mass.} Same as Figure \ref{fig:model_dep_Rd}, except linestyles indicate the amount of log-normal scatter, with $\sigmaN=0.1$, 0.2, and 0.3. We assume an intermediate case for high-mass SFE, $f_{\ast,\mathrm{hi}} \propto M_h^{-0.4}$, and a shallow limit for the dust scale length, $R_d \propto M_h^{1/3}$. The overall effect is a net blueward bias, as objects up-scattered into brighter magnitude bins always outnumber the typical object in that bin. Each model here takes $f_{\ast,10}=0.05$, $\Mpeak=2.8\times 10^{11} \ \Msun$, $\alphalo=2/3$, $\alphahi=-0.4$, $R_{d,10} = 2.4 \ \mathrm{kpc}$, and $R_d \propto M_h^{1/3}$.}
\label{fig:model_dep_scatter}
\end{center}
\end{figure*}

We show this effect in Figure \ref{fig:model_dep_scatter}. Due to the net bias toward bluer colors, models with more $N_d$ scatter can accommodate shallower relationships between $R_d$ and $M_h$, hence our adoption of the $R_d \propto M_h^{1/3}$ limit for each model in Figure \ref{fig:model_dep_scatter}. Without scatter, $\beta(\MUV)$ is much too sharp, as shown also in the dashed lines of Figure \ref{fig:model_dep_Rd}, but non-zero scatter curbs this behavior. There is tension between UVLFs and $\MUV-\beta$, which varies as a function of redshift, though this tension can be alleviated by slighly generalizing the $R_d$ parameterization and calibrating the model properly via multi-dimensional fitting, as we describe in the next sub-section.

\subsection{Model Calibration} \label{sec:calib}
In order to properly calibrate the model and quantify degeneracies between star formation and dust parameters, we perform a multi-dimensional Markov Chain Monte-Carlo (MCMC) fit to the $z \sim 4,6$, and 8 UVLFs from \citet{Bouwens2015} and $z \sim 4$ and 6 $\MUV$-$\beta$ relations from \citet{Bouwens2014} using \textsc{emcee}\footnote{\href{https://emcee.readthedocs.io/en/stable/}{https://emcee.readthedocs.io/en/stable/}} \citep{ForemanMackey2013}. 

\defcitealias{Bouwens2015}{B15}

We note before moving on to the results of this calibration that fitting to the \citetalias{Bouwens2014} empirical $\beta$ fits is more efficient computationally than, e.g., fitting to the \citet{Finkelstein2012} UV slopes determined via SED fitting. In order to compare fairly with the \citet{Finkelstein2012} measurements one needs higher wavelength resolution in order to adequately sample the spectra of objects within the \citet{Calzetti1994} spectral windows, of which there are 10, in contrast to the usual $\sim 2-5$ HST filters used in the \citetalias{Bouwens2014} analysis. Our approach scales as the number of wavelengths over which to perform spectral synthesis, making the empirical approach a more efficient option. We compare our best-fitting models to the \citet{Finkelstein2012} (hereafter F12) results shortly.

\defcitealias{Finkelstein2012}{F12}

The simplest model we explore has a total of six free parameters: the typical four parameters needed to describe a double power-law SFE ($f_{\ast,10}$, $\Mpeak$, $\alpha_{\mathrm{lo}}$, $\alpha_{\mathrm{hi}}$), and two parameters for the dust scale length ($R_{d,10}$, $\alpha_d$). We do not allow any of these parameters to evolve with cosmic time.

\begin{figure*}
\begin{center}
\includegraphics[width=0.98\textwidth]{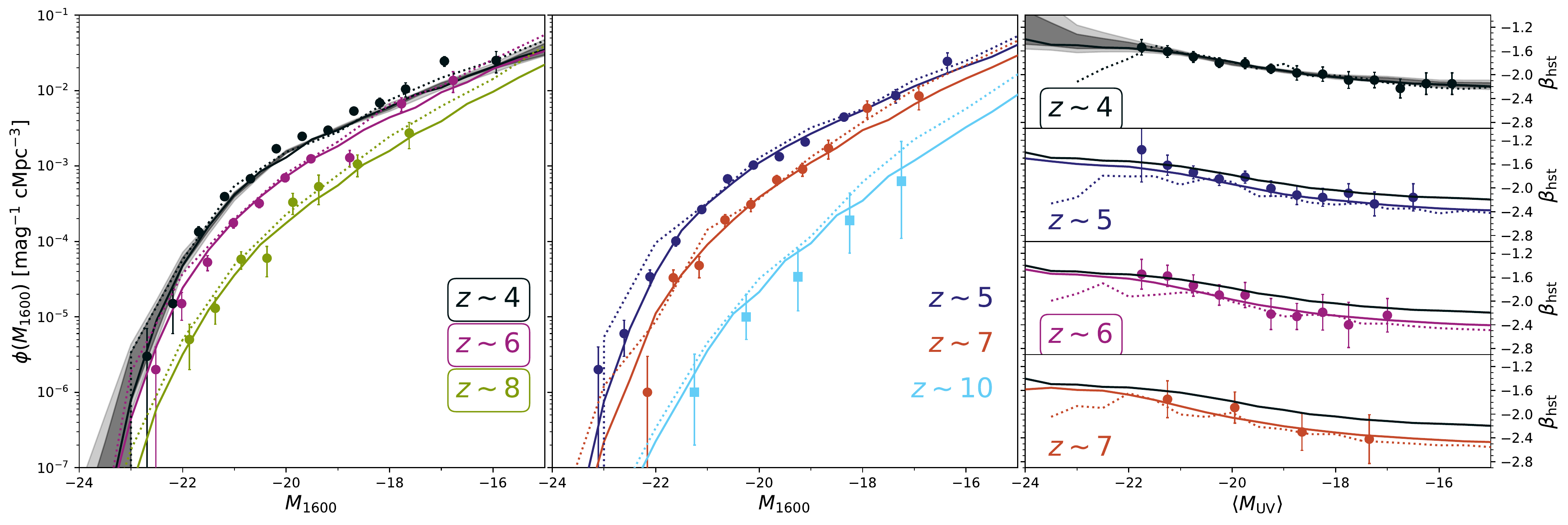}
\includegraphics[width=0.98\textwidth]{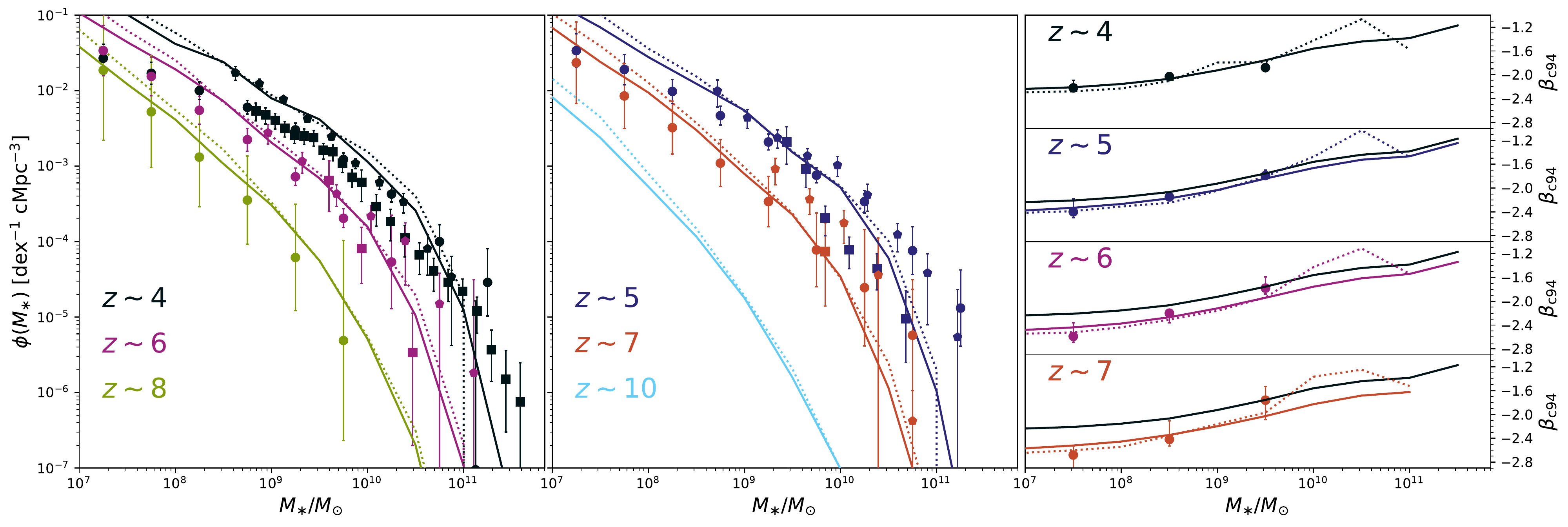}
\caption{{\bf Evolution of UVLF, SMF, and CMDs.} \textit{Top:} Rest UV information only, including UVLFs at $z \sim 4,6,$ and 8 (left), $z\sim 5,7,$ and 10 (center), and $\MUV-\beta$ relation from $4 \leq z \leq 7$. Data shown include UVLFs from \citet{Bouwens2015} ($4 \lesssim z \lesssim 8$) and \citet{Oesch2018} ($z \sim 10$), and $\MUV$-$\beta$ from \citetalias{Bouwens2014} (right). \text{Bottom:} Predictions in terms of stellar masses, rather than $\MUV$, including SMFs at $z \sim 4,6,$ and 8 (left), $z\sim 5,7,$ and 10 (center), and $\Mstell-\beta$ relations at $4 \leq z \leq 7$ (right). Data shown include SMFs from \citet{Song2016} (circles), \citet{Stefanon2017} (squares), and \citet{Duncan2014} (pentagons), and $\Mstell-\beta$ from \citet{Finkelstein2012}. Note that the \citet{Stefanon2017} and \citet{Duncan2014} stellar masses have been shifted by 0.25 dex to convert from a Chabrier to Salpeter IMF, and the $z\sim 4$ UV colours are repeated in each panel. Measurements used in the calibration are indicated with boxes around the corresponding redshifts.}
\label{fig:recon}
\end{center}
\end{figure*}

This simple, redshift-independent but halo mass-dependent model for star formation and dust obscuration agrees reasonably well with observations as shown in Figure \ref{fig:recon} (dotted lines). However, due to the tight link between star formation and dust production, the decline in the SFE at high-mass needed to match the steepness of the $z \sim 4$ UVLF has two unfortunate side-effects: (i) a turn-over in the $\MUV$-$\beta$ relation, and (ii) decline in the bright-end of the SMF much steeper than suggested by constraints from \citet{Song2016} and \citet{Stefanon2017} (see dotted lines in bottom row in Fig. \ref{fig:recon}).

We employ two strategies to remedy this problem in all that follows. First, we impose a prior requiring $\beta$ to be a monontonic function of $\MUV$ over the range of magnitudes probed by observations (including UVLFs and CMDs), which either eliminates a turn-over in $\MUV$-$\beta$ entirely or pushes it to slightly brighter objects, helping to reduce the disagreement between the bright-end of the UVLF and SMF. Second, we introduce an additional degree of freedom in our parameterization of $R_d$, allowing it to be a double power-law in $M_h$ rather than a single unbroken power-law\footnote{Physically, this could be a signature of morphological changes occurring at high mass. Alternatively, because $R_d$ is degnerate with $f_d$ and $\kappa$, it could be an indicator of changes in how dust is produced and/or destroyed in high-mass galaxies.}. With this parameterization, as the SFE declines at high-mass to match the steepness of the UVLF, the dust scale length can become shallower to ensure that $\beta$ continues to rise. This solution is amenable to shallower SFE curves at high-mass, resulting in better agreement with SMFs at the bright-end as well. Along with the standard four parameters for the SFE, this results in a total of 9 free parameters, which we calibrate via fitting to the \citetalias{Bouwens2014} $\MUV-\beta$ relation at $z=4$ and 6, and UVLFs from \citetalias{Bouwens2015} at $z\sim 4, 6$ and $8$. Best-fitting values of the model parameters and their uncertainties are summarized in Table \ref{tab:parameters}, with a subset of the posterior distributions shown in Figure \ref{fig:dpl_pdfs}.

In Figure \ref{fig:recon}, we show the rest UV calibration of this final model (top) and its predictions for the SMF as $\Mstell$-$\beta$ relation (bottom) at all $4 \lesssim z \lesssim 10$ (solid lines). The top row of Fig. \ref{fig:recon} is not terribly surprising, as much of the data shown is used in the calibration. Most noteworthy in this context is the evolution in the $\MUV$-$\beta$ relation, which arises despite the assumption that the production rate, opacity, and scale length of dust are constant in time. This evolution arises due to evolution in the typical stellar age, but also because specific star formation rates rise rapidly with redshift, which is a generic prediction of most models \citep[e.g.,][]{Behroozi2013b,Dayal2013}. In other words, part of the evolution in $\MUV$-$\beta$ is due to evolution in $\MUV$ (at fixed stellar or halo mass) alone, with the rest arising due to the bluer colors typical of increasingly young stellar populations at high redshift (see \S\ref{sec:evolution} for more discussion). From the bottom row of Figure \ref{fig:recon}, it is clear that none of our models predict a SMF as shallow as \citet{Song2016} at the faint-end. While jointly fitting UVLFs and SMFs is one potential solution to this problem, we have opted for a pure rest-UV-based approach in order to avoid complicating the calibration procedure further. Setting a prior on $d\beta/d\MUV < 0$ is a simple way to avoid tension between model and data at the bright-end, while remaining agnostic about issues at the faint-end of the SMF.

\begin{figure}
\begin{center}
\includegraphics[width=0.49\textwidth]{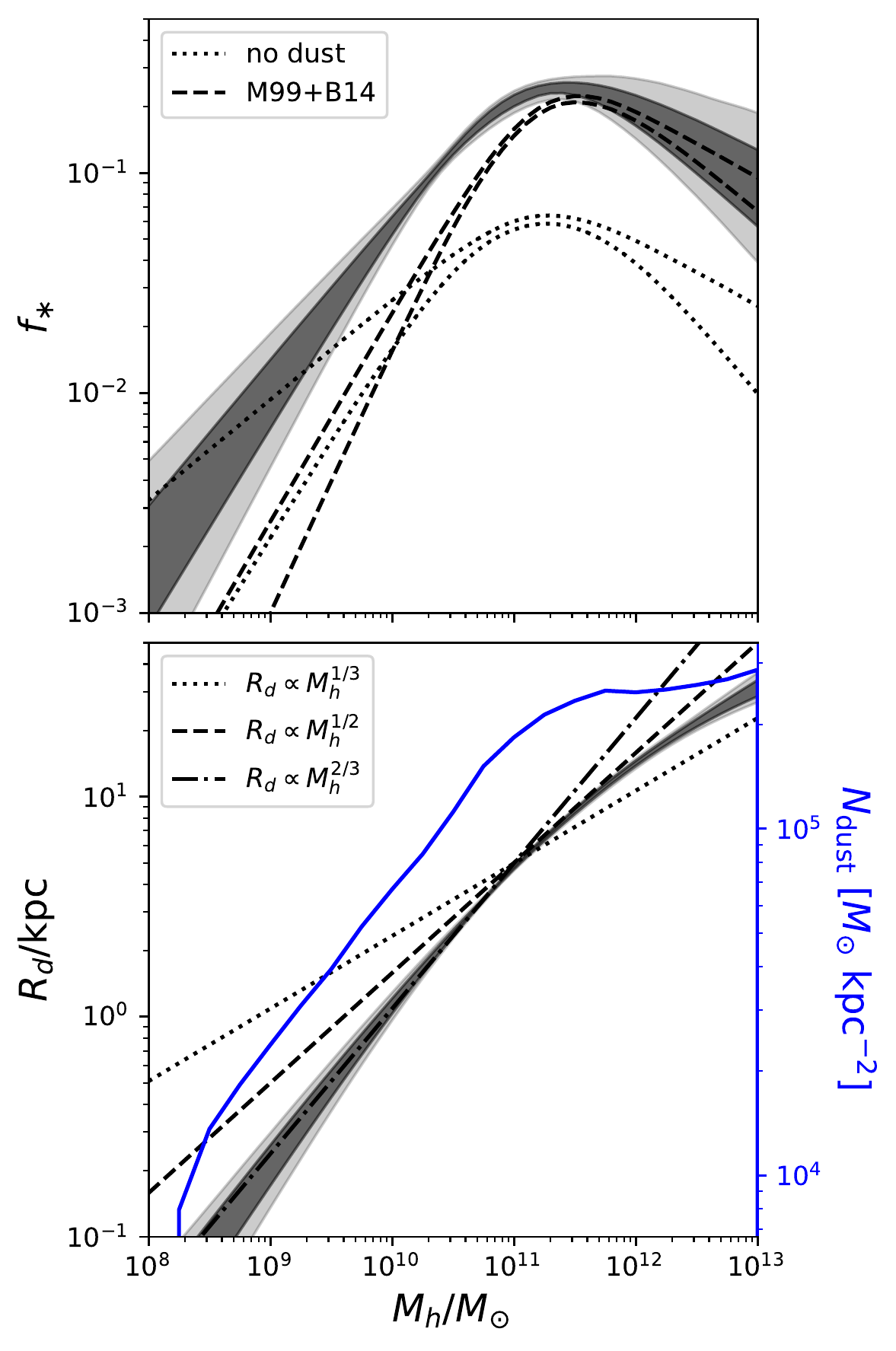}
\caption{{\bf Reconstructed star formation efficiency and dust scale length.} \textit{Top:} Filled contours indicate 68 and 95\% confidence intervals on the SFE using the 9 parameter model in this work, compared to a model with no dust calibrated to $z\sim 6$ (dotted), and a model using the common IRX-$\beta$-based approach (\citetalias{Meurer1999}+\citetalias{Bouwens2014}; dashed) calibrated with UVLFs at $z\sim 4$ and 6. Note that our recovered low-mass slope, $\alphalo \simeq 0.8$, is consistent both with the predictions of energy-regulated feedback models ($\alphalo = 2/3$) and steeper $\alphalo = 1$ scenarios \citep[e.g.,][]{Tacchella2018} at the $\sim 2\sigma$ level. \textit{Bottom:} Recovered dust scale length (black) and corresponding dust column density for best-fitting model (blue; right axis).}
\label{fig:recon_sfe}
\end{center}
\end{figure}

In Figure \ref{fig:recon_sfe}, we show the key ingredients of our model as recovered via MCMC fitting. First, in the top panel we show the SFE (filled gray contours) compared to a dust-free solution (dotted) and a solution obtained via the \citetalias{Meurer1999}+\citetalias{Bouwens2014} approach \footnote{Note that these functions are very similar to those presented in \citet{Mirocha2017} and \citet{Mirocha2019}, except we have replaced the \citet{Sheth2001} mass function with the \citet{Tinker2010} mass function and re-run the fit to be consistent with the current work.} (dashed). The \citetalias{Meurer1999}+\citetalias{Bouwens2014} approach converts intrinsic UV magnitudes to observed UV magnitudes by solving for the extinction $\AUV$ required to simultaneously satisfy the link between $\AUV$ and $\beta$ put forth in \citetalias{Meurer1999} and the connection between observed $\MUV$ and $\beta$ reported in \citetalias{Bouwens2014}. As expected, the treatment of dust affects both the normalization and shape of the SFE as a function of $M_h$, with offsets of a factor of $\sim 2-3$ near the peak. The posterior distribution for the component parameters, as well as the reconstructed SFRD, are included in Appendix \ref{sec:sfe_posterior}. In the bottom panel, we show the recovered dust scale length with pure power-laws included to guide the eye. The departure from a pure power-law is subtle -- at high-mass, $M_h \gtrsim 10^{11} \ \Msun$, our solution roughly tracks the $R_d \propto \Rvir \propto M_h^{1/3}$ solution, while at lower mass a steeper relation is preferred. The blue line shows the corresponding dust column density for the best-fit model only (right axis).

\begin{table}
\begin{tabular}{ | l | l | l | }
\hline
parameter & recovery & prior range \\
\hline
$\log_{10}(f_{\ast,\mathrm{norm}})$ & $-1.26^{+0.061}_{-0.022}$ & (-3, 0)\\
$\log_{10}(M_{\ast,\mathrm{peak}} / M_{\odot})$ & $11.16^{+0.165}_{-0.186}$ & (9, 13)\\
$\alpha_{\ast,\mathrm{lo}}$ & $0.80^{+0.100}_{-0.143}$  & (0, 1.5) \\
$\alpha_{\ast,\mathrm{hi}}$ & $-0.53^{+0.244}_{-0.020}$ & (-1, 1.5)  \\
\hline
$R_{d,10} / \mathrm{kpc}$ & $1.12^{+0.096}_{-0.083}$ & (0.01, 10)\\
$\alpha_{d,\mathrm{lo}}$ & $0.09^{+0.217}_{-0.068}$ & (0, 2) \\
$\alpha_{d,\mathrm{hi}}$ & $0.69^{+0.159}_{-0.023}$ & (-1, 2) \\
$\log_{10}(M_{d,\mathrm{peak}} / M_{\odot})$ & $12.01^{+0.314}_{-1.154}$ & (9, 13) \\
$\sigma_{\log_{10} N_d}$ & $< 0.039$ & (0, 0.6)\\
\end{tabular}
\caption{{\bf Marginalized 68\% constraints on the parameters of our fiducial model.} The first block of four parameters are those describing the SFE (see Eq. \ref{eq:sfe_dpl} and \S\ref{sec:stars}), while the next five parameters describe the dust scale length and scatter in dust column density (see Eq. \ref{eq:Rd_dpl} and \S\ref{sec:dust}). Fits were performed using broad uninformative priors on each parameter, as listed in the final column.}
\label{tab:parameters}
\end{table}

Finally, in Figure \ref{fig:dpl_pdfs}, we show a subset of the posterior distribution. From the left-most panel, we see the degeneracy between components of the double power-law $R_d$ model. Solutions favoring a single, unbroken power-law would track the dotted line, but clearly such solutions are not preferred by our fits. The maximum likelihohod model has a steep slope at high-mass, $\alphahiD \sim 0.1$, and shallower slope at low-mass, $\alphaloD \sim 0.7$, with a change in slope occurring at $10^{11} \lesssim M_{d,\mathrm{peak}} / \Msun \lesssim 10^{12}$ (panel b). There is of course a mild degeneracy  between the dust scale length and high-mass SFE slope (panel c), with $\alphahi \simeq -0.5$. The dust scale length and scatter have no significant degeneracy (panel d). Finally, in panel (e), we see that the halo mass of the break in the dust scale length is poorly constrained, and though $M_{d,\mathrm{peak}} > M_{\ast,\mathrm{peak}}$ is preferred, the 68\% contours are consistent with occurring at $M_{d,\mathrm{peak}} = \Mpeak$. A triangle plot of the SFE parameters is included in Appendix \ref{sec:sfe_posterior} compared to the results of simpler models published in previous studies.

\begin{figure*}
\begin{center}
\includegraphics[width=0.98\textwidth]{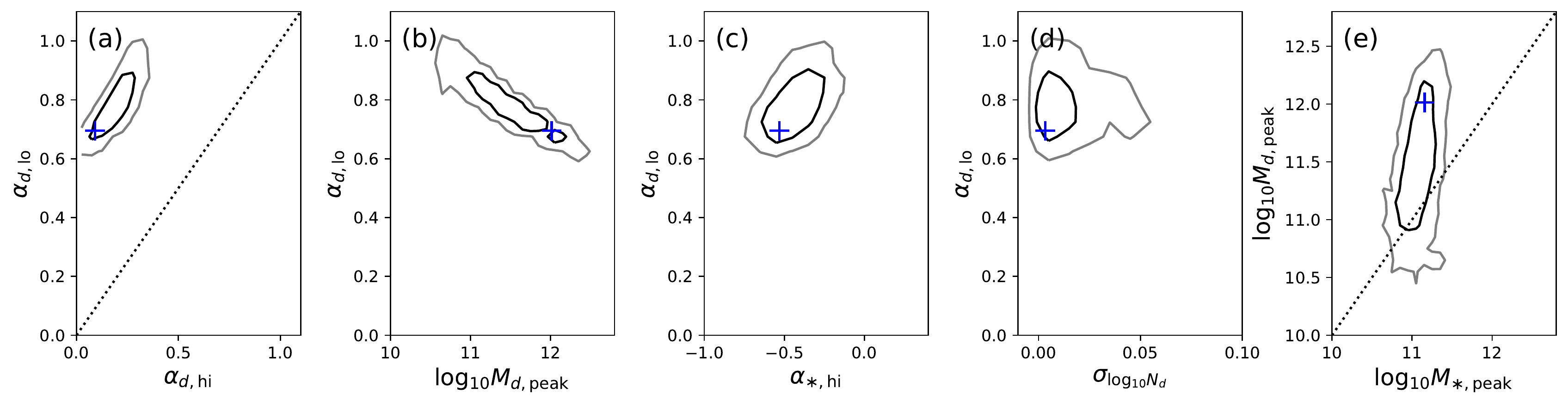}
\caption{{\bf Subset of posterior distributions for our fiducial 9-parameter model.} Black contours show 68 and 95\% confidence regions, while blue crosses indicate the maximimum likelihood point in each plane. Dotted 1:1 lines are included in first and last panels to assess the degeneracy between each component of $R_d$, and whether the peak mass in SFE and $R_d$ are consistent. Best-fitting parameters are summarized in Table \ref{tab:parameters}.}
\label{fig:dpl_pdfs}
\end{center}
\end{figure*}

\subsection{Model Predictions} \label{sec:predictions}
The bottom row of Figure \ref{fig:recon} shows our model's predictions for the galaxy SMF and relation between $\Mstell$ and $\beta$. Agreement is reasonably good, though, as is the case for many models \citep[e.g.,][]{Tacchella2018}, the slope of the $z \sim 4-5$ SMF at low-mass is in considerable tension with observational constraints. Our predictions are closer to the \citet{Duncan2014} SMF measurements than \citet{Song2016}, with a steep slope continuing toward even low-mass, as suggested by the measurements of \citet{Bhatawdekar2019}. At the bright end, our predictions agree well with the \citet{Stefanon2017} constraints. Predictions for $\Mstell$-$\beta$ are in good agreement with the constraints from \citet{Finkelstein2012}.

\begin{figure}
\begin{center}
\includegraphics[width=0.49\textwidth]{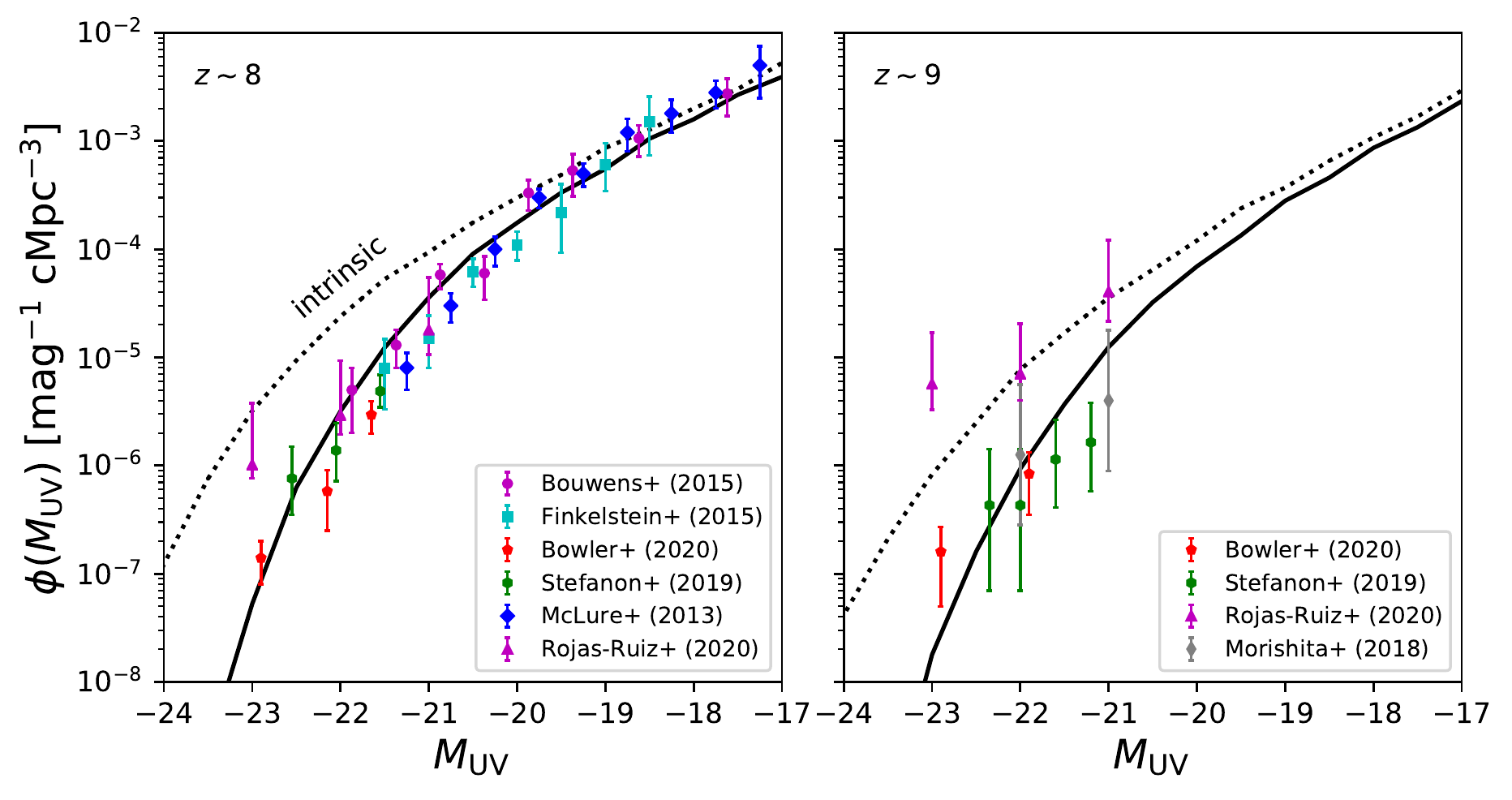}
\caption{{\bf A closer look at $z\sim 8-9$ UVLFs.} Dashed lines show our best-fit model before applying dust reddening, while the solid line is our best-fitting model including dust reddening and dimming. The relatively shallow bright-end reported recently by \citet{Bowler2020} is in reasonably good agreement with our predictions.}
\label{fig:z89}
\end{center}
\end{figure}

In Figure \ref{fig:z89}, we zoom-in on our predictions for the $z\sim 8$ and 9 UVLFs, given the continued progress in finding bright galaxies at these redshsifts \citep[e.g.,][]{Bowler2020,Morishita2018,Stefanon2019,McLeod2016,Livermore2018,RojasRuiz2020}. We show our predictions both for the observed UVLF (solid) and the intrinsic UVLF (dotted), i.e., the UVLF uncorrected for dust. We find that $z \sim 8-9$ galaxies are still reddened considerably at the bright end, by $\sim 1$ magnitude (at, e.g., $\MUVobs = -23$). It is difficult to compare fairly to the \citet{Bowler2020} measurements in full, as our models must match the systematically higher \citetalias{Bouwens2015} measurements by construction. The agreement at $\MUV \simeq -23$ may at least suggest that recent measurements do not necessarily indicate reduced dust content in bright galaxies at $z \sim 8-9$.

\begin{figure*}
\begin{center}
\includegraphics[width=0.98\textwidth]{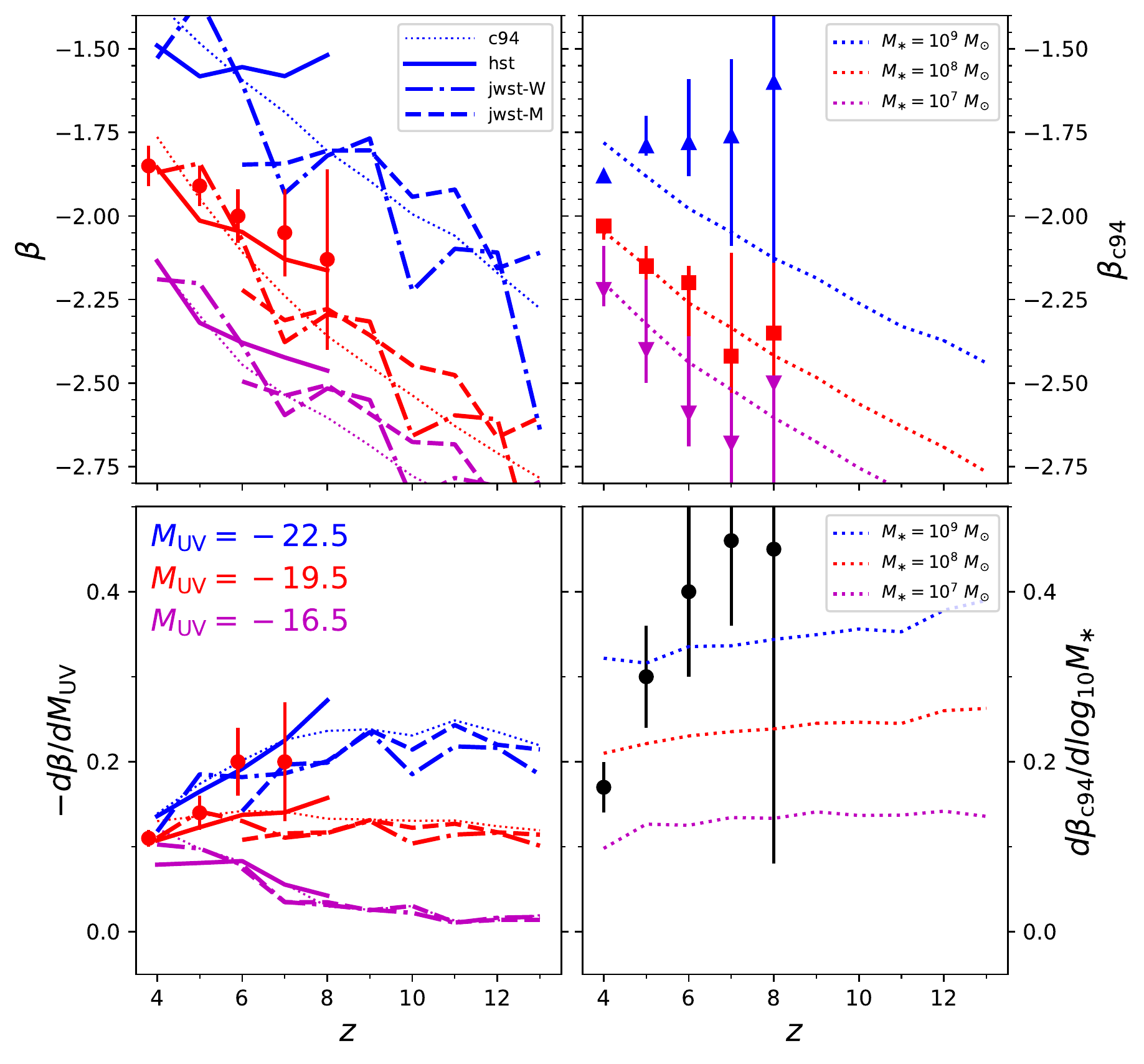}
\caption{{\bf Predictions for redshift evolution in UV colours at fixed $\MUV$ (left column; compared to \citetalias{Bouwens2014}) and fixed $M_{\ast}$ (right column; compared to \citetalias{Finkelstein2012}).} The top row shows the evolution of the UV slope in three different $\MUV$ (left) and $M_{\ast}$ (right) bins, while the bottom row illustrates evolution in the gradient of $\beta$ with respect to $\MUV$ (left) and $M_{\ast}$ (right). Continued evolution in $\beta$ at fixed $\MUV$ should be detected by JWST, with the most accurate recovery enabled by coverage in the NIRCAM medium filters (dashed). Pure wide-band photometry requires sampling the rest-UV spectrum outside the \citetalias{Calzetti1994} windows, resulting in a bias in $\beta$ estimates (dash-dotted). Evolution in $\beta$ at fixed $M_{\ast}$ is expected (top right), while little evolution in the gradient of the $\MUV-\beta$ and $M_{\ast}$-$\beta$ relations is expected at fixed $\MUV$ (bottom left) and $M_{\ast}$ (bottom right), respectively. Note that the \citetalias{Finkelstein2012} $M_{\ast}$-$\beta$ measurements are fit over $7.5 \leq \log_{10} \Mstell/\Msun \leq 9.5$ (black circles, bottom right panel), whereas we report the slope at three different $M_{\ast}$ values (see legend in top-right panel). Results in right column use $\beta$ as measured in the \citetalias{Calzetti1994} spectral windows, while $\beta$ values in left-column are computed using HST photometry (solid), NIRCAM wide (dash-dotted) and medium (dashed) filters, and \citetalias{Calzetti1994} windows (dotted). Refer to \S\ref{sec:synthobs} and Appendix \ref{sec:dust_biases} for more information about filter choices.}
\label{fig:jwst}
\end{center}
\end{figure*}

Because our model is fundamentally anchored to the evolution of dark matter halos, it is straightforward to make predictions for future UV colour measurements with JWST. We show these predictions in Figure \ref{fig:jwst}, including evolution in $\beta$ at various $\MUV$ (left) and $M_{\ast}$ (right). We expect the mild trend in $\beta(z;\MUV)$ observed thus far at $4 \lesssim z \lesssim 7$ to continue to higher redshift (top-left panel), as has been shown in other empirically-calibrated models \citep[e.g.,][]{Williams2018}. However, photometric measurements of $\beta$ generally do not recover the ``true'' UV colour evolution (computed using \citetalias{Calzetti1994} windows; dotted lines). For example, evolution in $\beta$ as computed with NIRCAM wide filters (dash-dotted) exhibits sharp features at redshifts where two-filter coverage requires excursions outside the \citetalias{Calzetti1994} range (see Figure \ref{fig:photometry} and Table \ref{tab:photometry}). The NIRCAM medium filters probe the underlying evolution more faithfully (dashed), at least at $z \gtrsim 7$, with only a slight red-ward bias, $\delta \beta \simeq 0.1$, as expected via photometry due to absorption lines in the stellar continuum. Further investigation into the evolution in the \textit{shape} of the $\MUV$-$\beta$ and $M_{\ast}$-$\beta$ relation (bottom row) seems potentially informative, as our models do not reflect the trends observed in \citetalias{Bouwens2014} and \citetalias{Finkelstein2012} in detail -- in fact, we predict little to no evolution in the $\beta$ gradients with respect to $\MUV$ (bottom left) or $\Mstell$ (bottom right). However, measurement uncertainties are large, so we have not investigated potential sources of this disagreement in detail at this stage.

\begin{figure}
\begin{center}
\includegraphics[width=0.49\textwidth]{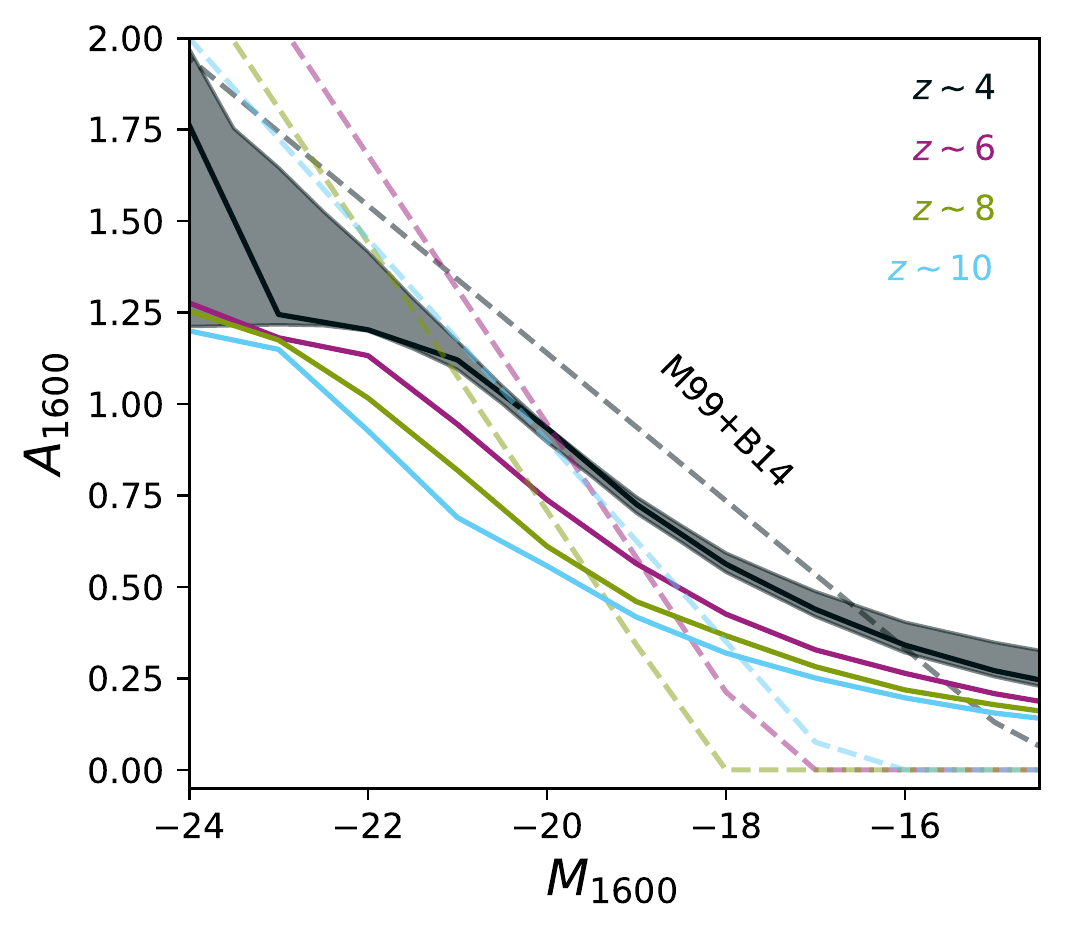}
\caption{{\bf UV extinction as a function of UV magnitude at $1600\angstrom$ in best-fitting model.} Dashed lines show the predictions from the \citetalias{Meurer1999} IRX-$\beta$ relation and \citetalias{Bouwens2014} $\MUV$-$\beta$ relation from $z \sim 4$ to $z \sim 10$, while our model predictions are shown as solid lines. Note that the relationship between $\AUV$ and $M_{\ast}$ is redshift-\textit{independent}, as we have assumed a constant dust yield per stellar mass and redshift-independent dust scale length, so we do not show it here. Gray shaded region indicates the 68\% confidence region for the reconstructed $z\sim 4$ $\AUV(\MUV)$ relation.} 
\label{fig:AUV}
\end{center}
\end{figure}

A key advantage of our approach is that we do not invoke an IRX-$\beta$ relationship to correct for dust, instead self-consistently solving for the UV luminosity and colors of high-$z$ galaxies with a semi-empirical dust model. As a result, the relationship between $\AUV$ and $\MUV$ is a prediction of our model, rather than an input. We show our recovered $\AUV(\MUV)$ curves in Figure \ref{fig:AUV} compared to the results obtained when assuming a \citet{Meurer1999} relation and the \citetalias{Bouwens2014} fits to $\MUV-\beta$. Our predicted $\AUV$ values are systematically lower than the \citetalias{Meurer1999}+\citetalias{Bouwens2014} approach for bright galaxies (dashed lines).

Finally, an interesting question in high-$z$ galaxy evolution is whether or not redshift evolution in $\MUV$-$\beta$ and Lyman-$\alpha$ emitter (LAE) fractions are related to the same underlying phenomenon. Evolution in both colors \citep[e.g.,][]{Finkelstein2012,Bouwens2014} and $\Lya$ emission \citep[e.g.,][]{Shapley2003,Pentericci2009,Verhamme2008,Stark2010,Hayes2011,Yang2017,Oyarzun2017} has been attributed to evolution in dust, but to our knowledge there has been no effort to connect these phenomena explicitly in a physical model. To explore this potential link, we make the simplifying assumption that any object with sufficiently low dust opacity will be a LAE, with the critical opacity left as a free parameter to be determined.

In a model with a 1:1 relationship between halo mass and dust column density, there will be a characteristic mass (or $\MUV$) at which galaxies become LAEs (assuming some equivalent width cut) -- this mass is set simply by the dust column density for which $\tau_{\mathrm{dust}} \sim 1$. Scatter in dust column density has an interesting side-effect in this context: the transition from objects that are optically thick to dust at $1600\angstrom$ is no longer a sharp function of halo mass and/or $\MUV$. In our framework, scatter in dust column density is degenerate with the dust scale length: an intrinscally shallow $R_d(M_h)$ relationship (and thus steep $\MUV$-$\beta$ relation) can be counteracted by scatter, and vice-versa (see \S\ref{sec:scatter}). So, though we cannot self-consistently predict the LAE fraction, we can explore different regions of the posterior distribution to see if the preferred values of $\sigmaN$ are preferred also by LAE measurements.

\begin{figure*}
\begin{center}
\includegraphics[width=0.98\textwidth]{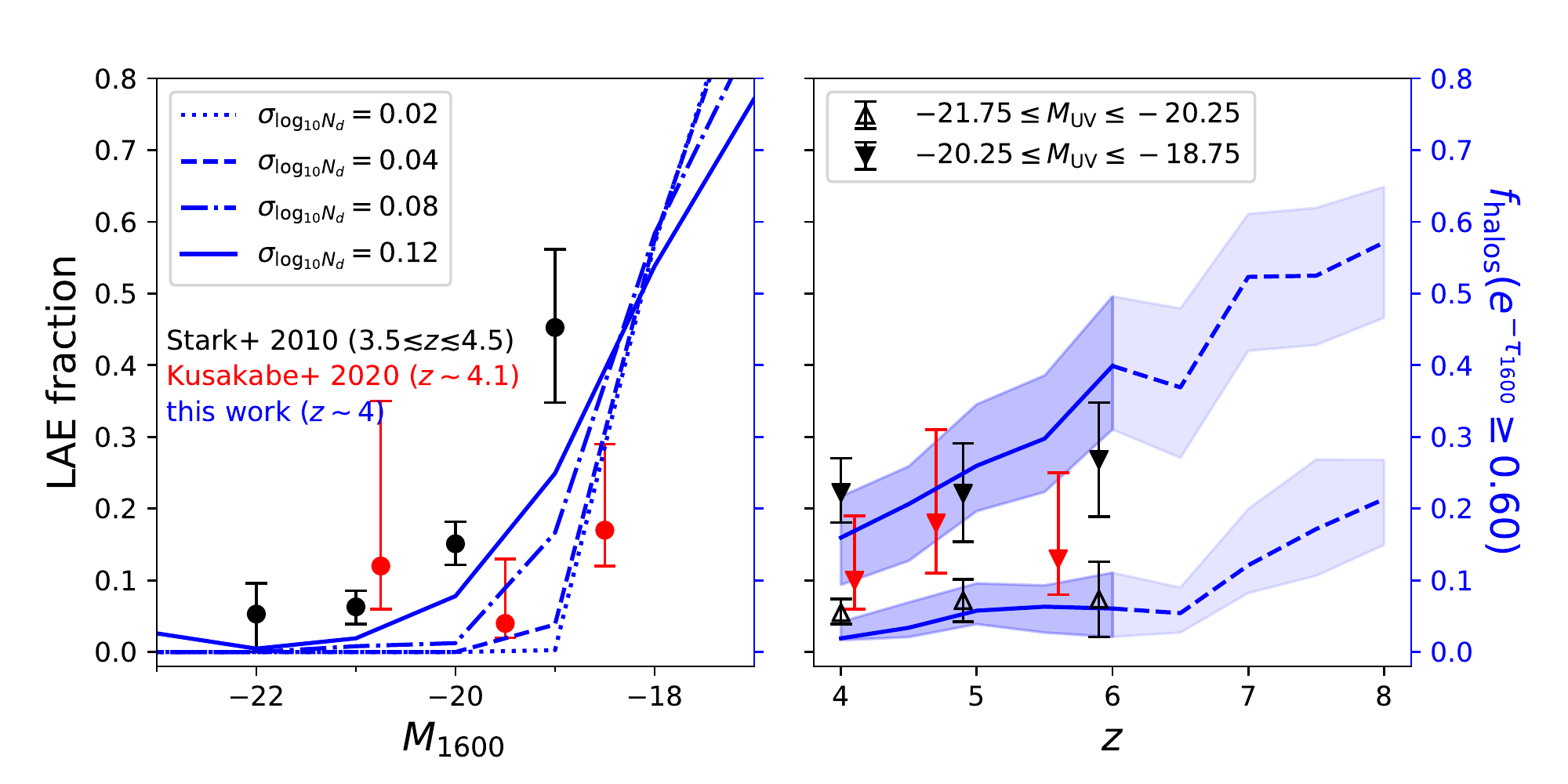}
\caption{{\bf Potential connection between $\MUV$-$\beta$ and LAEs.} \textit{Left:} LAE fractions (using equivalent width cut $\geq 55 \angstrom$) reported by \citet{Stark2010} from $3.5 \lesssim z \lesssim 4.5$ compared to fraction of objects with $1600\angstrom$ transmission in excess of $60 \pm 2.5$\% (right axis; blue) for four different levels of scatter in dust column density. \textit{Right}: Evolution in LAE emitter fractions with redshift from \citet{Stark2011}, separately for bright ($-21.75 \leq \MUV \leq -20.25$) and faint ($-20.25 \leq \MUV = -18.75$) objects. Shaded areas explore small deviations in the critical transmission ($\pm 0.025$). Model predicions at $z\gtrsim 6$ would be subject to the effects of reionization, hence the change in linestyle/shading in the right panel. In both panels, we also show the more recent measurements of \citet{Kusakabe2020} for comparison, which take  EW $\geq 55 \angstrom$ (both panels) and $-20.25 \leq \MUV \leq -18.75$ (right panel).}
\label{fig:LAEs}
\end{center}
\end{figure*}

As shown in Figure \ref{fig:LAEs}, the fraction of objects with high $1600\angstrom$ transmission ($e^{-\tau_{1600}} \geq 0.6 \pm 0.025$) looks remarkably similar to the LAE fraction, $\xLAE$, at $3.5 \lesssim z \lesssim 4.5$ as measured by \citet{Stark2010}, at least for $\sigmaN \gtrsim 0.12$ (solid lines). With less scatter, the fraction of objects with high UV transmission transitions more abruptly between zero and one (dotted curves). The redshift evolution of $f_{\mathrm{halos}}(>e^{-\tau_{1600}})$ in coarse $\MUV$ bins also agrees reasonably well with the redshift evolution measured by \citet{Stark2011}. The more recent \citet{Kusakabe2020} measurements are perhaps more accommodating of models with little scatter, $\sigmaN < 0.1$, as preferred in our fits, at least for $\MUV \gtrsim -20$. Of course, the caveat here is that the critical value of $e^{-\tau_{1600}} \geq 0.6 \pm 0.025$ was tuned by-eye until the normalization of models and measurements matched. Despite this, it is at least intriguing that $\sigmaN$ values permitted by $\MUV$-$\beta$ measurements generate reasonable LAE populations, and that relatively little scatter, $\sigmaN \simeq 0.1$, is needed to do so. Further exploration of this effect, e.g., how to accommodate larger values of $\sigmaN$, may thus be warranted.

\begin{figure}
\begin{center}
\includegraphics[width=0.49\textwidth]{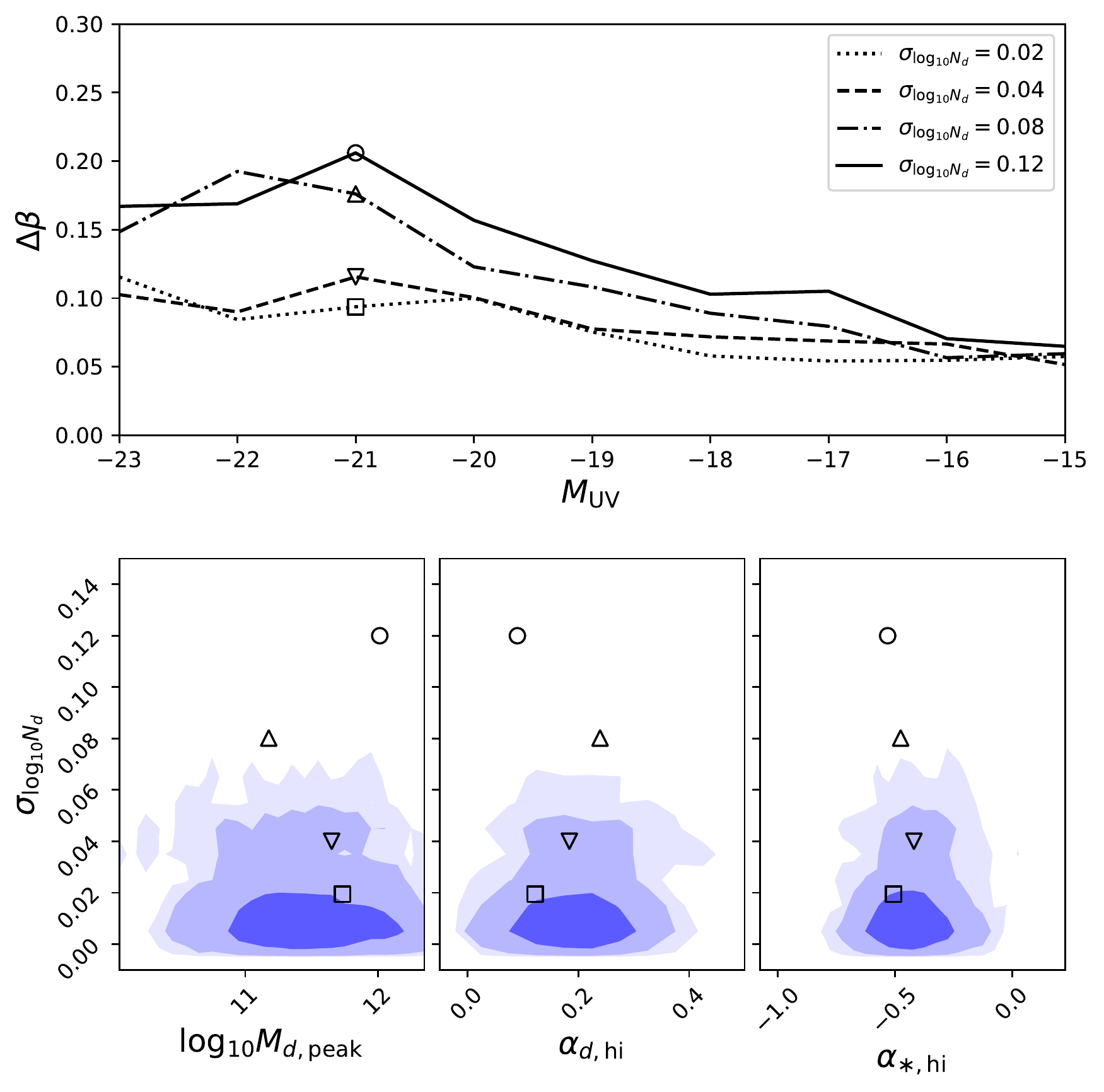}
\caption{{\bf Intrinsic scatter in $\beta(\MUV)$, $\Delta \beta$, at $z \sim 5$ (top) at various points drawn from the posterior (bottom).} Models shown are the exact same as those shown in Fig. \ref{fig:LAEs}, from negligible scatter (dotted) to $\sigmaN \simeq 0.12$ dex (solid). Recall from Fig. \ref{fig:LAEs} that $\sigmaN \gtrsim 0.1$ is generally preferred by LAE measurements. Scatter in SFR at fixed halo mass is responsible for the baseline $0.05 \lesssim \Delta \beta \lesssim 0.1$ scatter, in the limit of $\sigmaN \approx 0$. \textit{Bottom:} Location of models in most relevant dimensions of posterior distribution.}
\label{fig:scatter}
\end{center}
\end{figure}

Because the $\MUV$-$\beta$--LAE connection is largely an issue of scatter in dust column in our framework, we show also our model's predictions for the intrinsic scatter in $\beta$ at fixed $\MUV$, $\Delta \beta$. A larger amount of scatter in $N_d$ of course results in more scatter also in $\beta$, as we see in the top row of Fig. \ref{fig:scatter}. In order to ensure that variations in $\sigmaN$ would not worsen agreement with measured UVLFs and CMDs, we draw points directly from the posterior (indicated in bottom row of Fig. \ref{fig:scatter}), except for values $\sigmaN>0.1$, for which there are none. With $\sigmaN \sim 0.12$, our predictions come close to the empirical findings of \citet{Rogers2014}, who found evidence of steadily rising intrinsic scatter with increasing galaxy luminosity. The scatter in $\beta$ in our models is well-approximated as a Gaussian, in line with the assumptions of \citet{Rogers2014} and empirical findings of \citep{Castellano2012}.

\section{Discussion} \label{sec:discussion}
Our model, though simple, remedies potential inconsistencies in IRX-$\beta$ approaches while making testable predictions for upcoming observations. In this section we discuss the implications of the model, and assess the degree to which it is a useful conceptual framework for thinking about dust reddening in high-$z$ galaxies.

\subsection{Physical Interpretation of the Model} \label{sec:interpretation}
Taken at face value, our model predicts the UV luminosity and reddening of galaxies under the assumption that dust within galaxies is distributed uniformly in a sphere and is a source of attenuation only (i.e., no scattering), while star formation is centrally-concentrated (see Eq. \ref{eq:tau_d}). Only in this limit does the dust scale length uniquely determine both the dust density and the path length through dust to stellar sources. In reality, the distribution of dust in galaxies is unlikely to be so ideal, so we do not adhere strongly to this geometrical interpretation. Instead, we think of this model as a simple way to connect dust reddening to halo properties. In other words, we caution against over-interpreting the dust scale lengths we infer, and instead emphasize bulk properties like the dust mass, column density, UV luminosity, and colours. Because of this, it is perhaps more reasonable to refer to an \textit{effective} dust scale length or column density, i.e., that which is representative of the reddening over an entire galaxy, composed of many distinct star-forming regions and dust columns.

A key input to our model, aside from the dust scale length described above, is of course the dust production efficiency. We assume a constant dust yield $f_d = 0.4$ \citep{Dwek1998}, which fixes the dust-to-metal ratio (DTMR) in each of our model galaxies. Because we also assume a constant metal production efficiency, and instantaneous return model in which $\dot{M}_Z \propto \dot{M}_{\ast}$, the dust masses of galaxies in our model will always be a constant fraction of the stellar mass. In reality, the situation is likely more complicated. For example, the DTMR likely scales with halo mass in a non-trivial way depending on the interplay between dust production, destruction, and grain growth in the ISM. However, semi-analytic models including simple treatments of these processes generally predict variations over $7 \lesssim \log_{10} (M_{\ast}/M_{\odot}) \lesssim 12$ of only a factor of $\sim 2-3$, and perhaps $\sim 10$ in extreme cases \citep{Popping2017}. Given this rather shallow modulation of dust content with stellar mass, we do not expect our results to dramatically change upon generalizing the model.

State of the art numerical simulations generally do not include a network of processes specific to the formation and evolution of dust \citep[though see, e.g.,][]{McKinnon2019,Li2019}, and instead link the dust content of grid zones (or gas particles) directly to the local hydrogen column, metallicity, and potentially temperature \citep{Ma2019,Vogelsberger2020}. Because the birth clouds of stars are generally unresolved, subgrid additions \citep[following, e.g.,][]{Charlot2000} are often necessary. Though this approach to dust production is still idealized, such simulations can of course investigate the effects of highly inhomogeneous and anisotropic dust distributions, which we cannot. Predictions for the effective dust column density as a function of halo (or stellar) mass could provide important guidance for simpler, semi-analytic models like ours. Given that our model allows flexibility in $R_d$, rather than tying it to $\Rvir$, it would be particularly interesting to see the extent to which simulations predict strong mass or redshift evolution in $R_d$, $N_{d,\mathrm{eff}}$, and/or the degree to which dust traces gas mass at high redshift.

\subsection{Evolving Dust?} \label{sec:evolution}
Given that our model adopts a fixed dust yield and a time-independent dust opacity and scale length, all evolution in UV colours (at fixed $\MUV$) arises from evolution in components of the model unrelated to dust. As a result, any observed evolution in $\beta$ at fixed $\MUV$ need not be due to evolution in the properties of dust, at least in the limit in which the SFE and $R_d$ are universal. Evolution in UV colours occurs naturally in our model due to two independent effects: (i) objects of fixed $\MUV$ are hosted by smaller halos at higher redshift and thus have less dust than objects of the same $\MUV$ at lower redshift (sSFR grows rapidly with $z$; see \S\ref{sec:stars}), and (ii) the mean stellar age is simply younger at high redshift.

There is a significant body of work suggesting the need for evolution in the properties of dust with cosmic time, some of which are largely empirical\footnote{These inferences still require assumptions about, e.g., the temperature of dust, evolution of which could masquerade as evolution in dust content. Furthermore, dust may be multi-phase, complicating procedures based on a single temperature \citep{Liang2019}.} \citep{Reddy2010,Capak2015}, while others invoke evolving dust properties to reconcile galaxy formation models with observational constraints \citep{Guo2009,Somerville2012,Yung2019a,Qiu2019}. The latter admit to being \textit{ad hoc}: time-independent dust properties result in dramatic under-prediction of bright galaxies at high redshifts \citep[e.g.,][]{Somerville2012,Yung2019a}.

The results of numerical simulations are varied. \textit{Illustris} prefers redshift evolution in the dust opacity \citep{Vogelsberger2020}, while the \textsc{fire} and \textsc{croc} simulations introduce no such evolution \citep{Ma2019,Khakhaleva2016}, but still obtain luminosity functions that are roughly in agreement with observations. \citet{Ma2019} concluded that their simulations are consistent with evolution in dust properties, though such evolution would have to be geometrical in nature given their assumption of a constant dust-to-metal (DTM) ratio.

In this work, we find that evolution in UV colours and UV extinction (at fixed $\MUV$), which are often interpreted as signatures of evolving dust, arise naturally even for scenarios in which no dust (or even metallicity) evolution is allowed. The general need for evolving dust in some theoretical models is likely due to the shrinking sizes of high-$z$ galaxies at fixed mass. For example, many SAMs connects the dust scale length to the scale length of galaxy disks \citep{Somerville2012,Qiu2019}, which shrink rapidly in concert with the virial radii of their host dark matter halos, $\Rvir \propto M_h^{1/3} (1+z)^{-1}$. We are able to reproduce this effect: for example, if we force the relationship between dust scale length and halo masses to look like the relation between halo virial radii and mass ($R_d \propto M_h^{1/3}$, neglecting the redshift dependence), our $\MUV$-$\beta$ relations and UVLFs are much too steep (see dashed and dash-dotted curves in Fig. \ref{fig:model_dep_Rd}). However, because our model can adjust the scale length as a free parameter, and appeal to scatter in dust columns densities to shallow out intrinsically steep $\MUV$-$\beta$ relations, we are able to avoid redshift-dependent dust optical depths. The implicit prediction here is that if the scale length of dust does not perfectly track the scale length of gas in galaxy disks, the intrinsic properties of dust in galaxies need not evolve with time. This prediction is in principle testable, now that spatially resolved dust continuum maps can be obtained for high-$z$ galaxies using ALMA \citep[e.g.,][]{Gullberg2018}.

We note before moving on that there are other potential sources of mild evolution in the UV colours of galaxies that we have not included, such as metallicity evolution \citep[e.g.,][]{Wilkins2016}, which will of coursee affect the intrinsic spectrum of galaxies and potentially amplify the effects of scatter in dust column density. Furthermore, below $z \sim 6$, asymptotic giant branch stars become a relevant source of dust production that we have effectively neglected by assuming an instantaneous dust return model. We defer a detailed discussion of these effects to future work.

\subsection{Scatter in Dust Column}  \label{sec:intrinsicscatter}
In simulations and some semi-analytic models, scatter in the dust column density will inevitably arise due to viewing angle effects \citep{Yung2019a}. It is unclear if our treatment of $\sigmaN$ as a free parameter is acting to mimic these effects. Our model predicts scatter in $\beta$ (at fixed $\MUV$) at the level of $\Delta \beta \simeq 0.2$ at $\MUV \gtrsim -19$ (see Fig. \ref{fig:scatter}). This is comparable, though slightly lower, than the observed \citet{Rogers2014} trend and predictions from \citet{Yung2019a}. Given the importance of scatter in potentially setting both the shape of $\MUV$-$\beta$ and LAE fractions, future constraints on $\Delta \beta$ -- including its distribution function -- may serve as an important discriminator between models, and help determine if the $\MUV$-$\beta$/LAE connection explored here is at work in real galaxies.

\subsection{Connection with LAE fraction} \label{sec:LAEs}
We find that the $\MUV$-$\beta$ $\MUV$-$\xLAE$ relations can be explained by scatter in dust column density at fixed halo mass, $\sigmaN \gtrsim 0.1$, assuming that $1600\angstrom$ transmission is a reliable predictor of whether or not galaxies are strong LAEs. In this case, ``strong'' means equivalent widths of $\geq 55\angstrom$, so as to compare directly with the measurements of \citet{Stark2010}.

This binary model for the LAE fraction is much simpler than others in the literature. A common approach is to model the intrinsic Ly-$\alpha$ equivalent width (EW), which requires assumptions about the SFR, escape fraction, and kinematics of galaxies, and then apply an EW cut in order to compare with observations \citep[e.g.,][]{Dayal2008}. Our results suggest that, to zeroth order, the abundance of LAEs could be set by the dust column density PDF of galaxies, and that this PDF is also responsible (in part) for the shape of the $\MUV$-$\beta$ relation.

Though the potential connection between patchiness and $\Lya$ emission we explore in Fig. \ref{fig:LAEs} has been considered before \citep[e.g.,][]{Neufeld1991,Hansen2006,Finkelstein2008,Finkelstein2009}, to our knowledge, there have been no attempts to draw an explicit connection between the $\MUV$-$\beta$ and $\MUV$-$\xLAE$ in a forward model, despite the fact that both are often attributed to dust. The viability of scatter in $N_d(M_h)$ as the mechanism responsible for each trend is in principle testable, perhaps most clearly via measurements of the intrinsic scatter in $\MUV$-$\beta$ (see \S\ref{sec:intrinsicscatter}).

\subsection{Color Selection Criteria}
Given that the scatter we invoke is log-normal, it is possible that galaxies in our model will experience fluctuations in dust column substantial enough to migrate outside the typical high-$z$ galaxy color selection windows. However, for the small values $\sigmaN \lesssim 0.1$ preferred by our fits, this is a small effect. For example, using the \citetalias{Bouwens2014} color selection criteria, and assuming the \citet{Madau1995} model for IGM absorption, we find that only $\sim 1-3$\% of objects would be falsely excluded from $z \gtrsim 4$ samples due to the effects of scatter. Falsely-excluded galaxies are generally those with $\SFR \gtrsim 10^2 \ M_{\odot} \ \mathrm{yr}^{-1}$, with rest UV colours in HST bands of $i_{775}-J_{125} \gtrsim 1$ and $B_{435} - V_{606} \sim 3$, and can thus be confused with red elliptical galaxies at $z \sim 1$ \citep{Coleman1980}. This region of color-color space is sparsely populated observationally \citep[see, e.g., Fig. 3 of ][]{Bouwens2015}, but future detections in this space may warrant further attention. Most of these objects ($\gtrsim 90$\%) have $\MUV \lesssim -20$ and so are detectable for HST.

\subsection{Implications for JWST} \label{sec:jwst}
Our model will be readily testable with constraints on high-$z$ galaxy counts and colors from JWST. The most noticeable tension in the model is at the low-mass end of the $4 \lesssim z \lesssim 6$ stellar mass function, where current measurements diverge \citep[e.g.,][]{Song2016,Stefanon2017,Duncan2014}, We note that, while our predictions are consistent with steep low-mass slopes \citep[e.g.,][]{Duncan2014,Bhatawdekar2019}, such steep slopes are known to cause tension with local group constraints \citep{Graus2016}. Constraints on the stellar mass function should improve considerably with JWST given the substantial expansion of coverage in the infrared, which will much more fully probe the rest-optical emission of high-$z$ galaxies, and thus potentially resolve the current disagreement. Improved colour constraints, particularly for massive galaxies, would also improve the model calibration. However, given the small field of view of NIRCAM, such constraints may only be possible for shallow, wide area surveys.

The redshift evolution in $\beta(\MUV)$ predicted by our model is also readily testable with JWST. In principle, using the medium NIRCAM filters, rest-UV colours can be measured out to $z\sim 15$, provided there are galaxies bright enough to detect at such high redshifts. We find only minimal evolution in the \textit{shape} of $\beta(\MUV)$ and $\beta(M_{\ast})$. The medium filters are key to all future $z \gtrsim 8$ colour constraints, as the wide filters are cannot cleanly isolate the rest UV continuum generally used for estimating $\beta$ (see Fig. \ref{fig:photometry}).

Finally, testing the hypothesis that the scatter in dust column density drives both $\MUV$-$\beta$ and $\MUV$-$\xLAE$ will require improved constraints on the intrinsic scatter in $\beta(\MUV)$. We compare favorably, to the \citet{Rogers2014} estimates of scatter in $\beta(\MUV)$, in that the scatter rises monotonically for increasingly bright galaxies, and for $\sigmaN \sim 0.1$ yields $\Delta \beta \simeq 0.2$ at $\MUV \sim -21$, in agreement with \citet{Rogers2014} (though slightly lower). Larger values of $\sigmaN$, likely closer to $\sigmaN \approx 0.2$ would improve agreement, though our fits clearly prefer $\sigmaN \lesssim 0.1$. This could simply be a limitation of the model -- perhaps alternate parameterizations or additional flexibility would permit larger $\sigmaN$ values. We leave this as an avenue to pursue in future work.

\subsection{Implications for Cosmic SFRD \& Reionization}
Assumptions about dust necessarily impact the inferred cosmic star formation rate density (SFRD), and thus predictions for reionization. Naively, one might expect dust-free models to provide a floor in predictions for the SFRD, since any increase in the dust content of galaxies will require enhancements to the SFE in order to preserve agreement with observed UVLFs. However, the presence of dust can also modulate the inferred \textit{shape} of the SFE (see Fig. \ref{fig:recon_sfe}). As a result, the introduction of dust will shift the SFRD upward at late times, $z \lesssim 10$, but potentially result in little overall change in the SFRD at earlier times, when fainter halos (which form stars less efficiently) become the dominant source population. Indeed, we find a steeper slope $f_{\ast} \propto M_h^{2/3}$ (or steeper) here compared to the $f_{\ast} \propto M_h^{1/2}$ scaling reported in \citet{Mirocha2017}, the latter of which neglected dust (see Appendix \ref{sec:sfe_posterior}). As a result, systematic uncertainties in modeling dust -- at least in the framework presented here -- are unlikely to cause a substantially earlier start to reionization or reheating. Any evidence for efficient star formation at $z \gtrsim 10$ from, e.g., kinetic Sunyaev-Z'eldovich constraints from the CMB \citep[e.g.,][]{Miranda2017} or global 21-cm signal measurements \citep{Bowman2018}, is most likely indicative of star formation in halos below the atomic cooling threshold \citep{Mirocha2019,Mebane2020}. This statement may be subject to change in models in which the SFE of galaxies grows with redshift or changes shape in non-trivial ways, a possibility which we defer to future work.

\subsection{Implications for IRX-$\beta$} \label{sec:irxb}
A common approach to dust-correcting rest-UV measurement of high-$z$ galaxies is to invoke empirical correlations between infrared excesses and $\beta$ \citep{Meurer1999}. The IR excess can be related to UV attenuation under the assumption of a known intrinsic UV slope and dust opacity, making it possible to convert the observed $\MUV$ to an intrinsic UV magnitude, and thus SFR. The standard \citetalias{Meurer1999} approach assumes $\beta_0 = -2.23$, appropriate for constant star formation in the \textsc{starburst99} models. Our models use the \textsc{bpass} models \citep{Eldridge2009} and assume galaxy star formation histories are rising rapidly and have scatter, which modulates the input $\beta_0$ for each galaxy. The effects of breaking the assumptions made in \citetalias{Meurer1999} has been pointed out previously also by other authors \citep[e.g.,][]{Wilkins2013}.

Our forward model generally predicts \textit{less} UV attenuation at fixed $\MUV$ than the \citetalias{Meurer1999} relation, at least at the bright end, $\MUV \lesssim -18$, consistent with other studies \citep[see, e.g.,][]{Mancini2016}. This is likely a byproduct of our joint inference approach, as the $\MUV$-$\beta$ relation and UVLFs have competing requirements (see \S\ref{sec:basics} and Fig. \ref{fig:model_dep_sfe}). For the faint, $\MUV \gtrsim -18$ dust-poor galaxies, our model predicts more attenuation than \citetalias{Meurer1999}, which makes sense given that our model galaxies are intrinsically bluer than $\beta_{\mathrm{in}} = -2.23$.

\section{Conclusions}
We have presented a simple, but self-consistent model for dust reddening in high-$z$ galaxies that does not require assumptions about the IRX-$\beta$ relationship. Instead, we flexibly parameterize the dust column density of galaxies as a function of halo mass, and link dust production directly to star formation. Upon calibrating the model parameters via joint-fitting of high-$z$ UV luminosity function and colour-magnitude relation constraints, we find that:
\begin{itemize}
    \item Models without redshift-dependent dust properties still predict evolution in $\MUV$-$\beta$ given that stellar ages are declining and specific SFRs are rising with redshift. In other words, much of the evolution in $\MUV$-$\beta$ reflects evolution in the typical halo (or stellar) mass of galaxies in our model, and thus their integrated dust production. This result is conservative given our neglect of metallicity evolution, which is also expected to result in UV colour evolution at fixed stellar mass (see \S\ref{sec:evolution}; Figures \ref{fig:recon} and \ref{fig:jwst}).
    \item This lack of evolution is at odds with other models in the literature, which require \textit{ad hoc} redshift evolution in the dust opacity in order to prevent excessive reddening in high-$z$ galaxies. This need may be real: if the dust scale length is related to the scale length of galaxy disks, and thus dark matter halo virial radii, it will contract rapidly in both halo mass and redshift and cause reddening to increase as well. Our model assumes no redshift evolution in the dust scale length (at fixed $M_h$) and so avoids this effect. Observationally, constraining the effective dust scale length may be difficult, so guidance from simulations may offer important insights in this context (see \S\ref{sec:interpretation}).
	\item Scatter in the relationship between dust column density and halo mass can help accommodate dust scale lengths that track halo virial radii (see Fig. \ref{fig:model_dep_scatter}). Furthermore, for values of the log-normal scatter in effective dust column density at fixed halos mass, $\sigmaN \simeq 0.1$, the evolution in the abundance of galaxies with $1600\angstrom \geq 0.6$ resembles the evolution in the LAE population from $3 \lesssim z \lesssim 6$ (see Fig. \ref{fig:LAEs}). This could be an indicator that the shape of $\MUV$-$\beta$ and $\MUV$-$\xLAE$ are driven by the same phenomenon.
	\item Measurements of the intrinsic scatter, $\Delta \beta$, provide important constraints on this aspect of the model. An increased scatter in the dust column density increases the scatter in $\beta$ as well, with scatter continuing to grow in even brighter objects (see Fig. \ref{fig:scatter}).
	\item UV colours are expected to continue to evolve smoothly with redshift at fixed $\MUV$, and are in principle measurable with JWST out to $z \sim 15$ at all $\MUV \lesssim -17$ (see Fig. \ref{fig:jwst}). We predict little to no redshift evolution in the shape of the $\MUV$-$\beta$ and $\MUV$-$M_{\ast}$ relations.
\end{itemize}
The aforementioned results depend on the assumption of a time-independent SFE, which is common in the recent literature, but also on a universal $R_d(M_h)$ relation. Both assumptions are subject to change in simple models, which predict evolution in the SFE and $R_d(M_h)$, if indeed $R_d \propto \Rvir$. We explore the consequences of these assumptions in a forthcoming paper (Mirocha et al., in prep). \\

J.M. acknowledges stimulating conversations with Chris Willott, Nissim Kanekar, Louis Abramson, Adrian Liu, Alan Heavens, James Rhoads, Tracy Webb, and Steve Finkelstein, and the anonymous referee for comments that helped improve this paper. J.M. also acknowledges support through a CITA National Fellowship. C.M. acknowledges support through the NASA Hubble Fellowship grant HST-HF2-51413.001-A awarded by the Space Telescope Science Institute, which is operated by the Association of Universities for Research in Astronomy, Inc., for NASA, under contract NAS5-26555. Computations were made on the supercomputers Cedar (from Simon Fraser University and managed by Compute Canada) and Mammouth (from the Universit\'{e} de Sherbrooke and managed by Calcul Qu\'{e}bec and Compute Canada). The operation of these supercomputers is funded by the Canada Foundation for Innovation (CFI), the minist\`{e}re de l'\'{E}conomie, de la science et de l'innovation du Qu\'{e}bec (MESI) and the Fonds de recherche du Qu\'{e}bec - Nature et technologies (FRQ-NT).

\textit{Software:} numpy \citep{numpy}, scipy \citep{scipy}, matplotlib \citep{matplotlib}, h5py\footnote{\url{http://www.h5py.org/}}, and mpi4py \citep{mpi4py1}.

\textit{Data Availability:} The data underlying this article is available upon request, but can also be re-generated from scratch using the publicly available \textsc{ares} code.

\bibliography{references}
\bibliographystyle{mn2e_short}

\appendix

\section{Photometric Estimates of UV Slopes} \label{sec:dust_biases}
It is important to extract $\MUV$ and $\beta$ from theoretical models in an observationally-motivated way, given that biases (in $\beta$ especially) comparable to observed trends can arise if using an idealized approach.

\begin{figure}
\begin{center}
\includegraphics[width=0.49\textwidth]{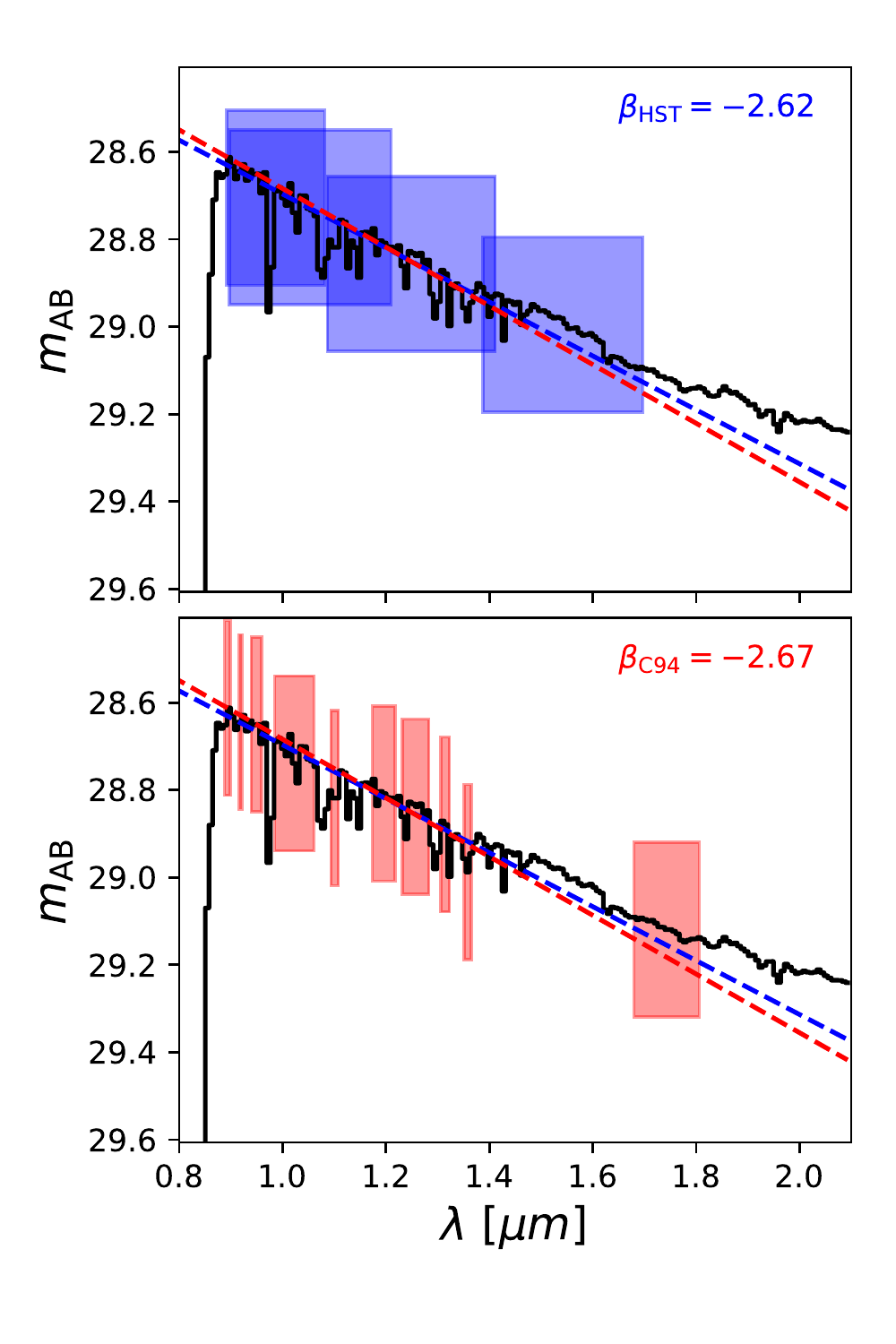}
\caption{{\bf Example extraction of $\beta$ from photometry.} The top panel shows the four HST filters used for galaxies at $z\sim 6$, while the bottom panel shows the placement of the \citet{Calzetti1994} spectral windows. Dashed lines show best-fitting power-laws using the respective photometric windows. Input spectrum corresponds to a dust-free galaxy from our model observed at $z=6$, with $M_{\ast}= 10^9 \ M_{\odot}$, $\dot{M}_{\ast} \simeq 3 \ M_{\odot} \ \mathrm{yr}^{-1}$, and $M_h \sim 10^{11} \ M_{\odot}$.}
\label{fig:spec_v_phot}
\end{center}
\end{figure}

\begin{figure}
\begin{center}
\includegraphics[width=0.49\textwidth]{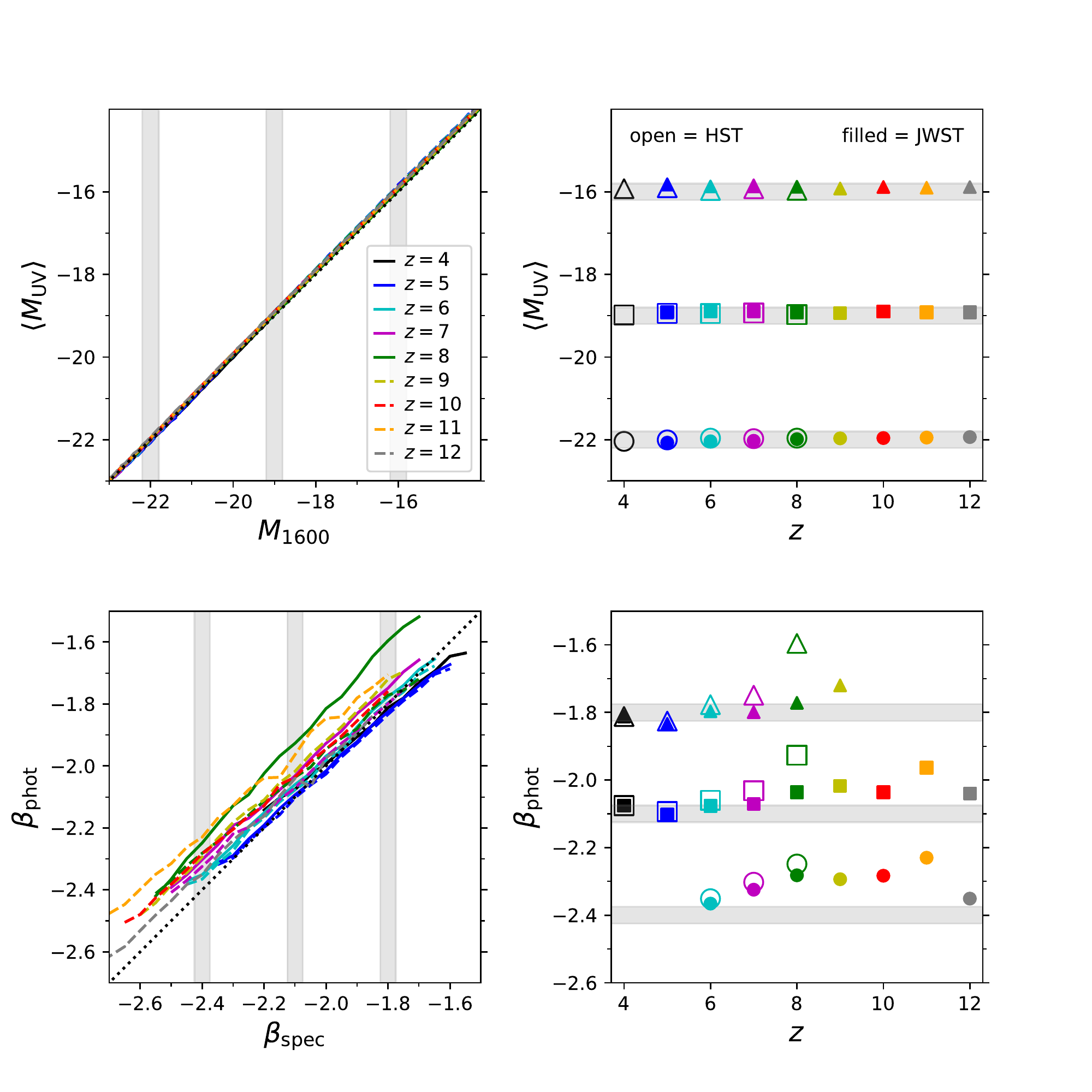}
\caption{{\bf Biases in photometrically-estimated UV magnitudes and colours.} \textit{Top:} UV magnitude, $\langle \MUV \rangle$, defined as geometric mean of available photometry, as a function of monochrotmatic UV magnitude, $M_{1600}$, at a series of redshifts(left), and as a function of redshift  holding $M_{1600}$ fixed at -22, -19, and -16 (right). \textit{Bottom:} UV colour estimated via power-law fits to available photometry, $\beta_{\mathrm{phot}}$, vs. the ``true'' UV color, $\beta_{\mathrm{spec}}$, computed via power-law fits to the forward-modeled galaxy SEDs in the \citet{Calzetti1994} spectral windows. We show $\beta_{\mathrm{phot}}$ as a function of $z$ in three fixed $\beta_{\mathrm{spec}}$ bins, $\beta_{\mathrm{spec}}$=-2.4, -2.1, and -1.9. Open symbols in panels in the right column indicate measurements performed only with HST WFC/WFC-3, while filled symbols use HST and JWST NIRCAM medium filters. See Table \ref{tab:photometry} for listing of filters used at each redshift.}
\label{fig:biases}
\end{center}
\end{figure}

In Figure \ref{fig:spec_v_phot}, we illustrate the difference between spectroscopic and photometric estimates of UV magnitudes and colours. In the former case, we assume $\MUV = M_{1600}$, measure $\beta$ as a power-law fit to intrinsic galaxy spectra through the \citetalias{Calzetti1994} windows. Our photometric estimates follow \citetalias{Bouwens2014}, who compute $\MUV$ as the geometric mean of all photometry, which we indicate with angular brackets, $\langle \MUV \rangle$. Similarly, $\beta$ is computed as a power-law fit through available photometry. In Figure \ref{fig:biases}, we show $\Delta \beta \simeq 0.05-0.2$ biases in the UV colour estimated with HST photometry (open sybmols), similar to what was found in \citet{Finkelstein2012} (see their Fig. 4), particularly at $z \gtrsim 6$. These biases will persist, even with JWST (see filled symbols in lower right panel of Fig. \ref{fig:biases}).

\begin{figure}
\begin{center}
\includegraphics[width=0.49\textwidth]{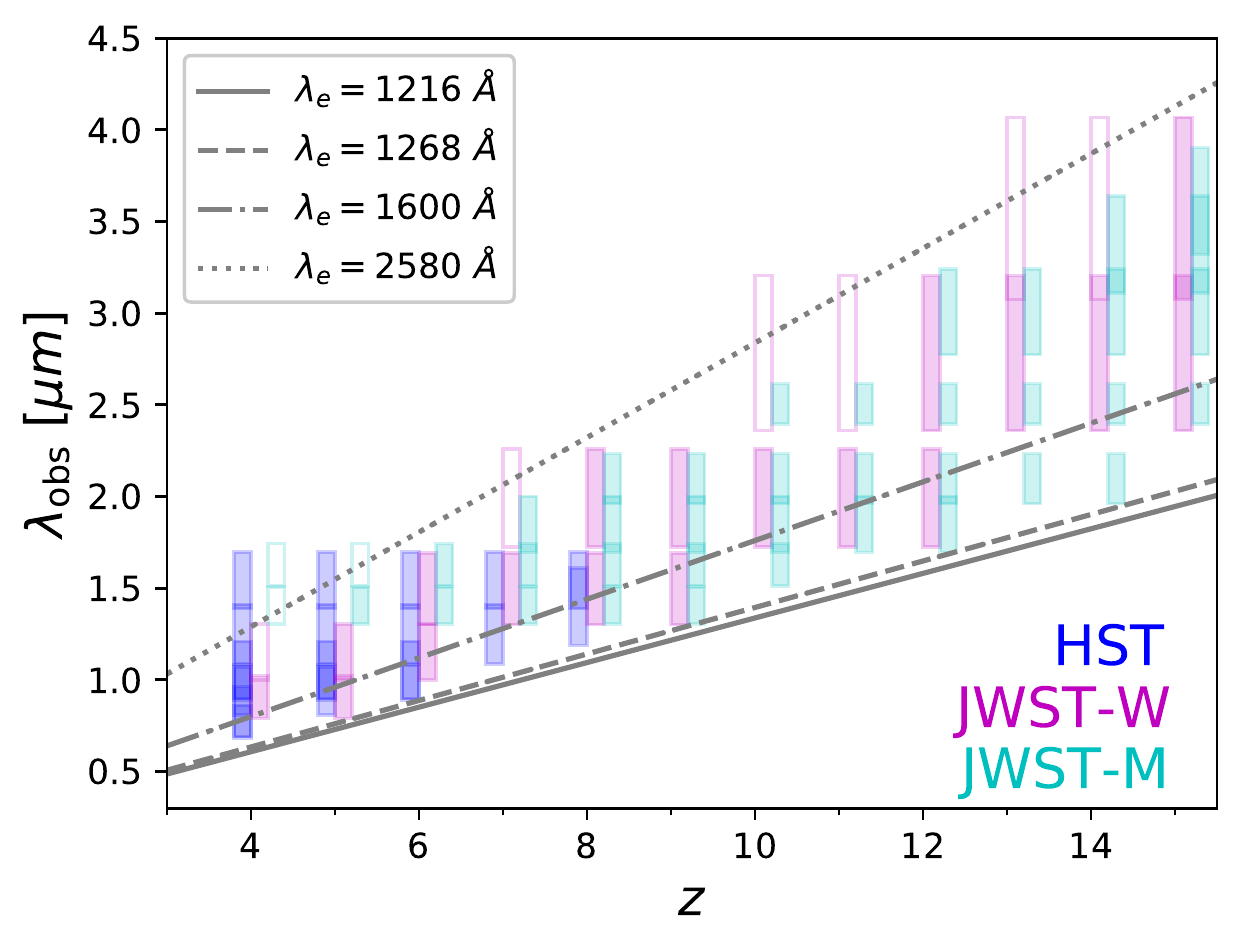}
\caption{{\bf Filter placement vs. galaxy redshift.} Blue boxes show filters used in HST analyses of \citetalias{Bouwens2014}, while filled magenta and cyan boxes show the NIRCAM wide (W) and medium (M) filters that lie completely within the \citetalias{Calzetti1994} windows. Open boxes are added to supplement the NIRCAM filter set at redshifts for which only one filter lies within \citetalias{Calzetti1994} windows. For reference, we show the observed wavelength of $\Lya$ (solid), the \citet{Calzetti1994} spectral range often used to compute $\beta$ (dashed, dotted), and $1600 \angstrom$, where magnitudes are often reported. See Table \ref{tab:photometry} for full filter listing. Note that boxes are offset horizontally for clarity, but placement is determined using integer redshifts only.}
\label{fig:photometry}
\end{center}
\end{figure}

From Figure \ref{fig:photometry} it is clear that at $z \gtrsim 6$, the rest $1268 \lesssim \lambda \lesssim 2580$ range is sampled more sparsely than at $z \lesssim 6$. This is particularly noticeable at $z \sim 7-8$ for HST, with coverage heavily weighted to the bluest part of the rest UV spectrum. Coverage at $9 \lesssim 11$ for JWST is also weighted to the bluest part of the rest-UV spectrum, which has many absorption features, hence the redward bias (bottom right panel of Fig. \ref{fig:biases}). Clearly, use of the NIRCAM medium filters will be required in order to get accurate colours at $z \gtrsim 8$. One could in principle use the wide filters at $z > 8$ also, though their increasing spectral width with wavelength corresponds to an increase in contamination from emission outside the \citetalias{Calzetti1994} spectral range (see unfilled boxes).

We include a full listing of the filters used as a function of redshift both for HST and the JWST wide and medium filters in Table \ref{tab:photometry}.

\begin{table*}
\begin{tabular}{| l | l | c | c | c | c | c | c | c | c | c | c }
\hline
filter & name & $z \sim 4$ & $z \sim 5$ & $z \sim 6$ & $z \sim 7$ & $z \sim 8$ & $z \sim 9$ & $z \sim 10$ & $z \sim 11$ & $z \sim 12$ & $z \sim 13$\\
\hline
F775W  & $i_{775}$  & \cmark & & & & & & & \\
F814W  & $I_{814}$  & \cmark & & & & & & & \\
F850LP & $z_{850}$  & \cmark & \cmark & & & & & & \\
F098M  & $Y_{098}$  & \cmark & \cmark & \cmark & \cmark & & & & & & \\
F105W  & $Y_{105}$  & \cmark & \cmark & \cmark & & & & & \\
F125W  & $J_{125}$  & \cmark & \cmark & \cmark & \cmark & & & & & & \\
F140W  & $JH_{140}$ & & & & & \cmark & \\
F160W  & $H_{160}$  & \cmark & \cmark & \cmark & \cmark & \cmark & & & & & \\
\hline
\hline
F090W  & n/a        & \cmark & \cmark & & & & & & & & \\
F115W  & n/a        & \xmark & \cmark & \cmark & & & & & & & \\
F150W  & n/a        & & & \cmark & \cmark & \cmark & \cmark & & & & \\
F200W  & n/a        & & & & \xmark & \cmark & \cmark & \cmark & \cmark & \cmark & \\
F277W  & n/a  & & & & & & & \xmark & \xmark & \cmark & \cmark \\
F356W  & n/a  & & & & & & & & & & \xmark \\
\hline
F140M  & n/a        & \cmark & \cmark & \cmark & \cmark & \cmark & \cmark & & & &  \\
F162M  & n/a        & \xmark & \xmark & \cmark & \cmark & \cmark & \cmark & \cmark & & & \\
F182M  & n/a        & & & & \cmark & \cmark & \cmark & \cmark & \cmark & \cmark &  \\
F210M  & n/a        & & & & & \cmark & \cmark & \cmark & \cmark & \cmark & \cmark \\
F250M  & n/a        & & & & & & & \cmark & \cmark & \cmark & \cmark \\
F300M  & n/a        & & & & & & & & & \cmark & \cmark \\
\hline
\hline
\end{tabular}
\caption{{\bf Filters used to estimate UV slope as a function of redshift.} First 8 rows indicate HST filters used by \citet{Bouwens2014} to estimate $\beta$, while bottom two blocks of filters demarcated by horizontal lines are the JWST NIRCAM wide and medium filters used in our analysis. We use \cmark\ symbols to indicate filters that lie within the spectral range of the \citetalias{Calzetti1994} filters, and \xmark\ symbols to indicate filters used out of necessity, when only one filter lies in the desired range. Use of undesirable filters at $z = 7, 10,$ and 13 are responsible for the sharp features in Fig. \ref{fig:jwst} when adopting the NIRCAM wide filters.}
\label{tab:photometry}
\end{table*}

\section{Updated SFE Constraints} \label{sec:sfe_posterior}
For completeness, here we present our new constraints on the SFRD and SFE parameters, and compare directly to two other common approaches: (i) use of the IRX-$\beta$ method of dust correction, or (ii) neglect of dust attenuation entirely.

\begin{figure}
\begin{center}
\includegraphics[width=0.49\textwidth]{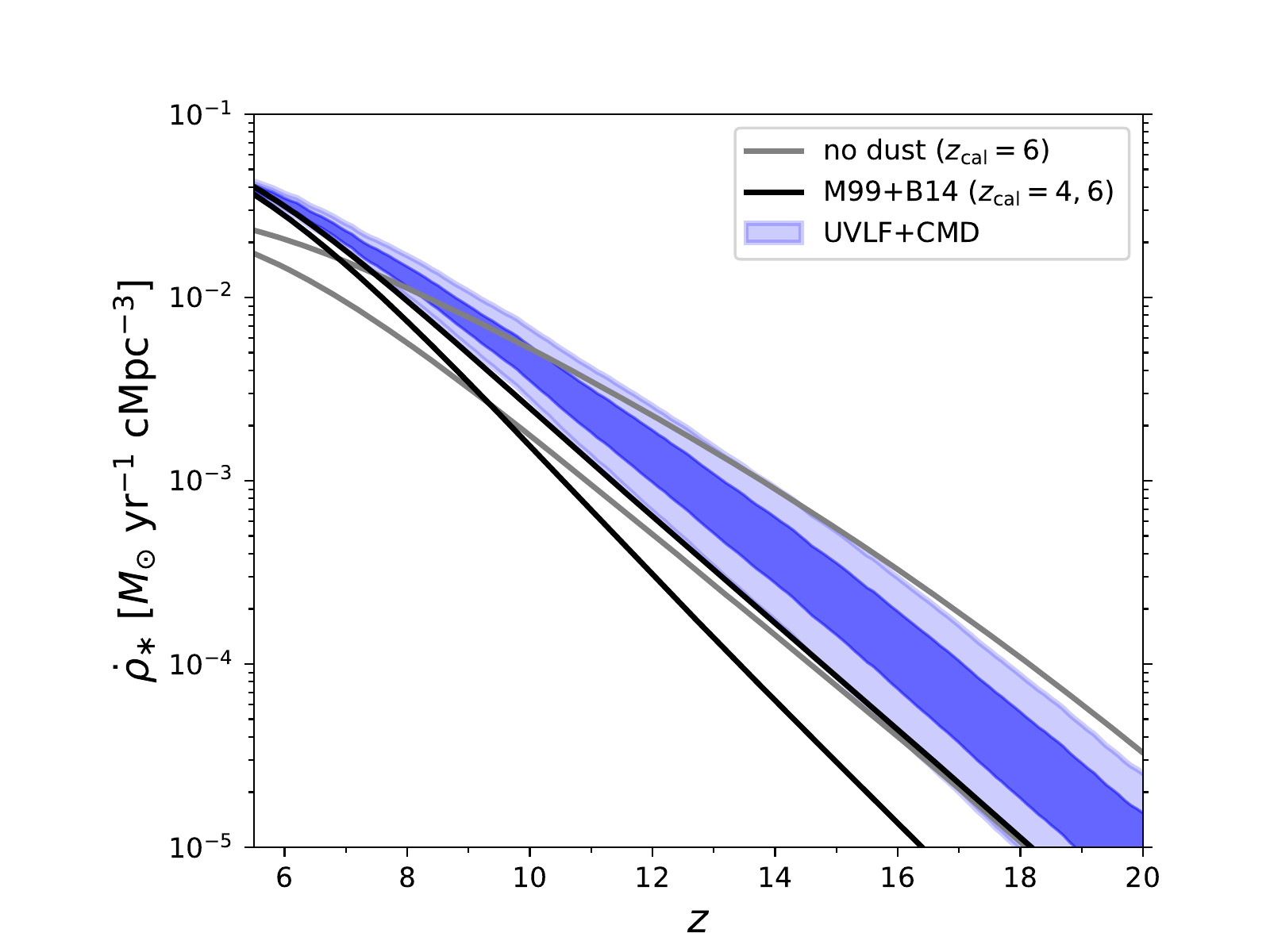}
\caption{{\bf Star formation rate density at high-$z$ predicted by our models.} Blue contours show 68\% and 95\% credibility regions for our new models presented in this work, while gray and black contours show the 68\% confidence regions obtained when neglecting dust and employing an IRX-$\beta$-based approach (described in text), respectively.}
\label{fig:sfrd}
\end{center}
\end{figure}

First, in Figure \ref{fig:sfrd}, we show the SFRD reconstructed from our fits compared to a dust-free model calibrated to $z \sim 6$ UVLFs from \citetalias{Bouwens2015} (gray contours), and a model calibrated to $z \sim 4$ and 6 UVLFs from \citetalias{Bouwens2015} using the common \citetalias{Meurer1999}+\citetalias{Bouwens2014} IRX-$\beta$-based approach (black contours). Our new models (blue) agree with the IRX-$\beta$ approach at $z \sim 6$, predicting $\dot{\rho}_{\ast} \simeq 4 \times 10^{-2} \ M_{\odot} \ \mathrm{yr}^{-1} \ \mathrm{cMpc}^{-3}$. At higher redshifts, the new models tend toward the dust-free $z \sim 6$ calibration because, while the inclusion of dust biases the normalization of the SFE high, it also biases the shape of the SFE toward steeper slopes. As a result, our new estimates of the $z \gtrsim 10$ SFRD are largely unchanged compared to previous work \citep{Mirocha2017}.

\begin{figure*}
\begin{center}
\includegraphics[width=0.98\textwidth]{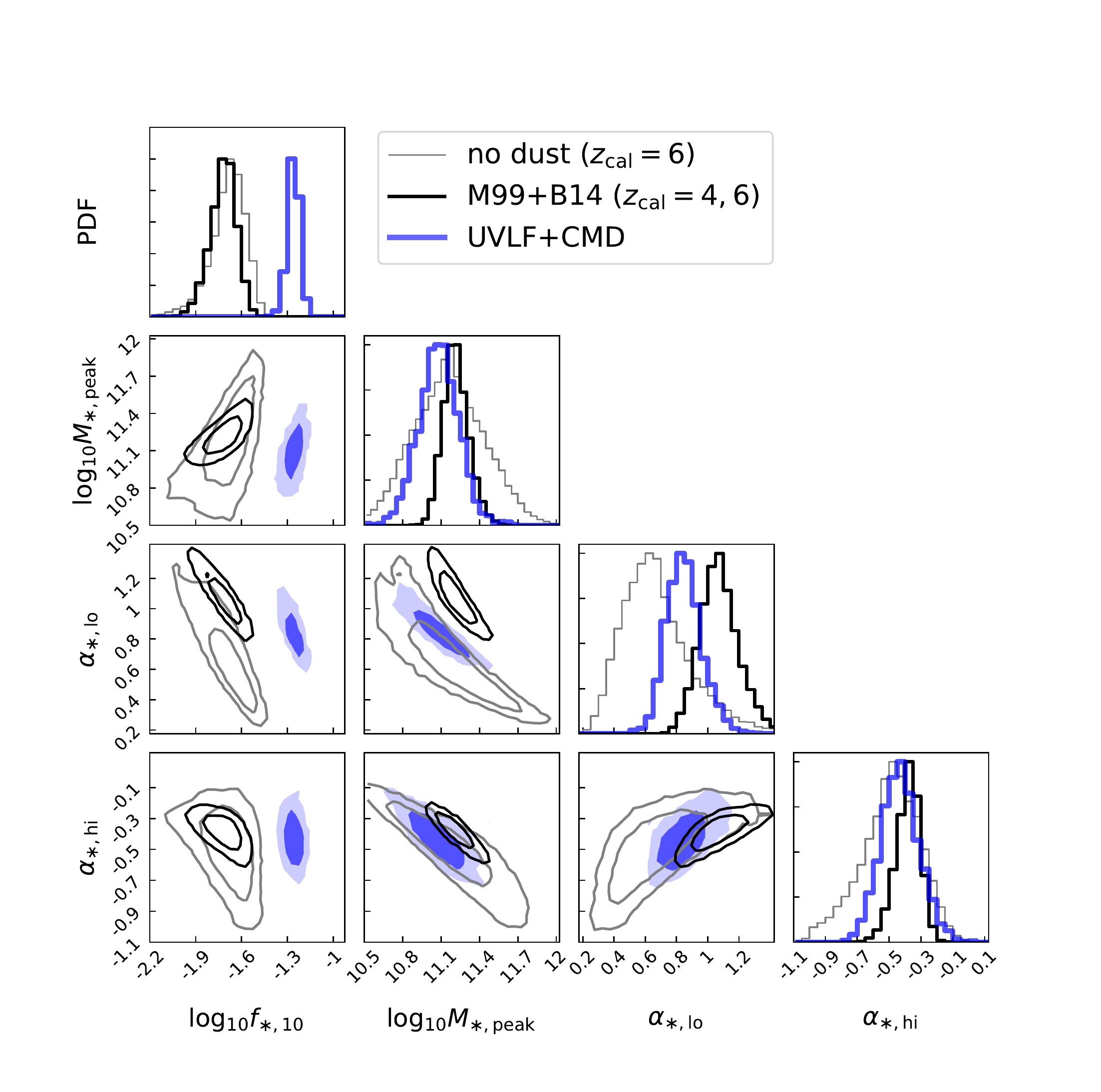}
\caption{{\bf Posterior distribution of SFE parameters for three different approaches to rest-UV inference.} Results using the new model presented in this work are shown in blue, while the dust-free and \citetalias{Meurer1999}+\citetalias{Bouwens2014} results are shown in gray and black, respectively. In each case, inner contours represent 68\% confidence regions, while outer contours indicate 95\%. Best fits and 1-$\sigma$ error-bars are presented in Table \ref{tab:parameters}.}
\label{fig:triangle_sfe}
\end{center}
\end{figure*}

In Figure \ref{fig:triangle_sfe}, we show the posterior distribution of the SFE parameters. Following the same color scheme as in Figure \ref{fig:sfrd}, we can quickly see that there are systematic differences when neglecting dust, particularly in the normalization of the SFE, which here we anchor to $10^{10} \ M_{\odot}$ halos (first column; $f_{\ast,10}$). Our new approach, while qualitatively similar to the \citetalias{Meurer1999}+\citetalias{Bouwens2014} contours, does exhibit some important differences. For example, the new models prefer shallower SFE slopes at low mass (third column; $\alpha_{\ast,\mathrm{lo}}$), which result in slightly higher SFRDs at $z \gtrsim 8$, as described above and shown in Figure \ref{fig:sfrd}.

\end{document}